\documentclass[
10pt,
showpacs,preprintnumbers,nofootinbib,
 amsmath,amssymb,
 aps,
prc,twocolumn,groupedaddress,superscriptaddress,
showkeys
]{revtex4-1}
\usepackage{graphicx}
\usepackage{dcolumn}
\usepackage{bm}
\usepackage[colorlinks=true,urlcolor=blue,citecolor=blue,breaklinks=true]{hyperref}
\usepackage{epsfig}

\newcommand{\leftsub}[2]{{\vphantom{#2}}_{#1}{#2}}

\begin{document}


\title{Three-cluster dynamics within an {\em ab initio} framework}

\author{Sofia Quaglioni}
 \email{quaglioni1@llnl.gov}
 \affiliation{Lawrence Livermore National Laboratory, P.O. Box 808, L-414, Livermore, California 94551, USA}
\author{Carolina Romero-Redondo}
 \email{cromeroredondo@triumf.ca}
\affiliation{TRIUMF, 4004 Wesbrook Mall, Vancouver, British Columbia, V6T 2A3, Canada}
 \author{Petr Navr\'atil}
 \email{navratil@triumf.ca}
 \affiliation{TRIUMF, 4004 Wesbrook Mall, Vancouver, British Columbia, V6T 2A3, Canada}

\date{\today}

\begin{abstract}
We introduce a fully antisymmetrized 
treatment of three-cluster dynamics within the {\em ab initio} framework of the no-core shell model/resonating-group method (NCSM/RGM). Energy-independent non-local interactions among the three nuclear fragments 
are obtained from realistic nucleon-nucleon interactions and consistent {\em ab initio} many-body wave functions of the clusters. The three-cluster Schr\"odinger equation is solved with bound-state boundary conditions by means of the hyperspherical-harmonic method on a Lagrange mesh. We discuss the formalism in detail and give algebraic expressions for systems of two single nucleons plus a nucleus. Using a soft similarity-renormalization-group evolved chiral nucleon-nucleon potential, we apply the method to an $^4{\rm He}$+$n$+$n$ description of $^6{\rm He}$ and compare the results to experiment and to a six-body diagonalization of the Hamiltonian performed within the harmonic-oscillator expansions of the NCSM. Differences between the two calculations provide a measure of core ($^4{\rm He}$) polarization effects. 
\end{abstract}
\maketitle


\section{\label{sec:introduction} Introduction}
In nuclear physics, {\em ab initio} approaches seek to solve the many-body Schr\"odinger equation in terms of constituent protons and neutrons interacting through nucleon-nucleon ($NN$) and three-nucleon ($3N$) forces that yield a high-precision fit of two- and three-body data.
Their aim is twofold: 
firstly, to help unfold the true nature of the force among nucleons; and secondly, to arrive at a fundamental understanding of nuclei and their role in the universe.

In the three- and four-nucleon systems, where a numerically exact solution of the quantum-mechanical problem for both negative~\cite{PhysRevC.64.044001} and positive energies~\cite{PhysRevC.84.054010} is now possible, this goal has been largely achieved.  
For heavier systems, {\em ab initio} calculations have been mostly confined to the description of the bound-state properties of stable nuclei, but are now starting to be extended to dynamical processes between nuclei. The Green's Function Monte Carlo method has been used to describe the elastic scattering of neutrons on $^4$He~\cite{PhysRevLett.99.022502} and to compute asymptotic normalization coefficients~\cite{PhysRevC.83.041001} and nuclear widths~\cite{PhysRevC.86.044330}. Loosely bound and unbound nuclear states have been addressed within the Coupled Cluster technique~\cite{Hagen2007169,PhysRevLett.104.182501} by using a Berggren basis and this method has been recently applied to compute elastic proton scattering on $^{40}$Ca~\cite{PhysRevC.86.021602}. 
 
An {\em ab initio} framework that promises to provide a unified treatment of a wide range of nuclear phenomena (well-bound states, loosely-bound and unbound exotic nuclei, scattering and reaction observables) is the no-core shell model with continuum (NCSMC)~\cite{PhysRevLett.110.022505,PhysRevC.87.034326}. 
Here, the nuclear many-body states are seen as superimpositions of continuous $(A-a,a)$ binary-cluster wave functions in the spirit of the resonating group method (RGM)~\cite{RGM,Tang1978167,Fliessbach198284,RGM3,Lovas1998265,PhysRevC.77.044002} and square-integrable eigenstates of the $A$-nucleon system, in which each cluster of nucleons and the compound nuclear states are obtained within the {\em ab inito} no-core shell model (NCSM)~\cite{PhysRevLett.84.5728,PhysRevC.62.054311}. So far, we have laid the foundations of the NCSMC by developing the formalism to compute nucleon-nucleus collisions and applying it to the description of the unbound $^7$He nucleus. However, expansions on the NCSM/RGM portion of the basis~\cite{PhysRevLett.101.092501,PhysRevC.79.044606} have been already successfully used to describe nucleon~\cite{PhysRevC.82.034609} and deuteron~\cite{PhysRevC.83.044609} scattering on light nuclei and achieve the first {\em ab initio} description of the $^7$Be$(p,\gamma)^8$B radiative capture~\cite{Navratil2011379} and the $^3$H$(d,n)^4$He and $^3$He$(d,p)^4$He fusion rates~\cite{PhysRevLett.108.042503}, based on realistic $NN$ interactions. 
Work is currently under way to incorporate the $3N$ force into this binary-reaction formalism and to attain the description of deuteron-nucleus scattering and transfer reactions within the NCSMC approach. 

Achieving an {\em ab initio} treatment of three-cluster dynamics is another important stepping stone towards gaining a basic understanding of nuclei and their reactions. To cite a few instances, important nuclear fusion processes such as the $^3$H$(^3$H$, 2n)^4$He or $^3$He$(^3$He$, 2p)^4$He reactions are characterized by three-body final states. In addition, only with an approach capable of accounting for three-cluster configurations can one obtain 
an accurate description of Borromean nuclei, ternary systems of two nucleons orbiting around a tightly bound core whose components are not bound in pairs. Finally, three-body configurations can be necessary even at very low energy to achieve a proper treatment of polarization and virtual excitations of breakup channels in reactions with weakly-bound projectiles such as the deuteron.  

Microscopic three-cluster models, where all nucleons are taken into account and the Pauli principle is treated exactly, have been used for some time, particularly in combination with the hyperspherical formalism for the solution of the dynamic equations~\cite{Filippov01091996,PhysRevC.63.034606,PhysRevC.63.034607,Korennov2004249,PhysRevC.80.044310,PhysRevC.85.034318}. However, they have two main limitations: the use of central 
$NN$ potentials with state-dependent parameters adjusted to reproduce the binding energy of the system under study, occasionally complemented with a spin-orbit interaction; and a simplified description of the internal structure of the clusters, which are in most cases described by $s$-shell wave functions.
In this paper, we report on an extension of the NCSM/RGM formalism  to treat the dynamics among three nuclei made of fully antisymmetrized interacting nucleons. The solution of the three-cluster Schr\"odinger equation is obtained by means of hyperspherical harmonic (HH) expansions on a Lagrange mesh~\cite{Descouvemont:2003ys,Descouvemont:2005rc}. In addition, we present the first $^4$He+$n$+$n$ investigation of the ground state (g.s.)\ of the $^6$He nucleus 
based on
a $NN$ potential that yields a high-precision fit of the $NN$ phase shifts and {\em ab initio} four-body wave functions for the $^4$He cluster obtained consistently from the same Hamiltonian. In particular, we employ a similarity-renormalization-group (SRG)~\cite{PhysRevC.75.061001,PhysRevC.77.064003} evolved chiral N$^3$LO $NN$~\cite{N3LO} potential.  For this first application, 
we include only the g.s.\ of the $^4$He cluster and estimate the importance of the core polarization by comparing the results obtained with six-body NCSM diagonalizations of the adopted Hamiltonian. The inclusion of excited states of $^4$He to describe such effects is hard and not very efficient within the NCSM/RGM approach. On the other hand, core-polarization effects will be easily accounted for once the present formalism will be embedded within the NCSMC framework. 
 
The paper is organized as follows. In section~\ref{sec:formalism} we define the microscopic three-cluster problem, present a brief overview of the HH functions and their application within the $R$-matrix method on Lagrange mesh for the solution of the three-body bound-state problem, and introduce in detail the three-cluster NCSM/RGM formalism. In particular, in Sec.~\ref{subsec:RGMkernels} we present algebraic expressions for systems of two single nucleons plus a nucleus. Results for the g.s.\ of the $^6$He Borromean nucleus
are presented in Sec.~\ref{sec:applications}, where we discuss calculations performed by solving the $^4$He(g.s.)+$n$+$n$ NCSM/RGM equations and compare them with a diagonalization of the Hamiltonian in the six-body NCSM model space. Conclusions and outlook are given in Sec.~\ref{sec:conclusions}. Finally, additional details on the formalism are presented in the Appendix.  

\section{\label{sec:formalism} Formalism}
\subsection{Microscopic three-cluster problem}
\label{subsec:microscopic}
\begin{figure}[t]
\includegraphics[width=60mm]{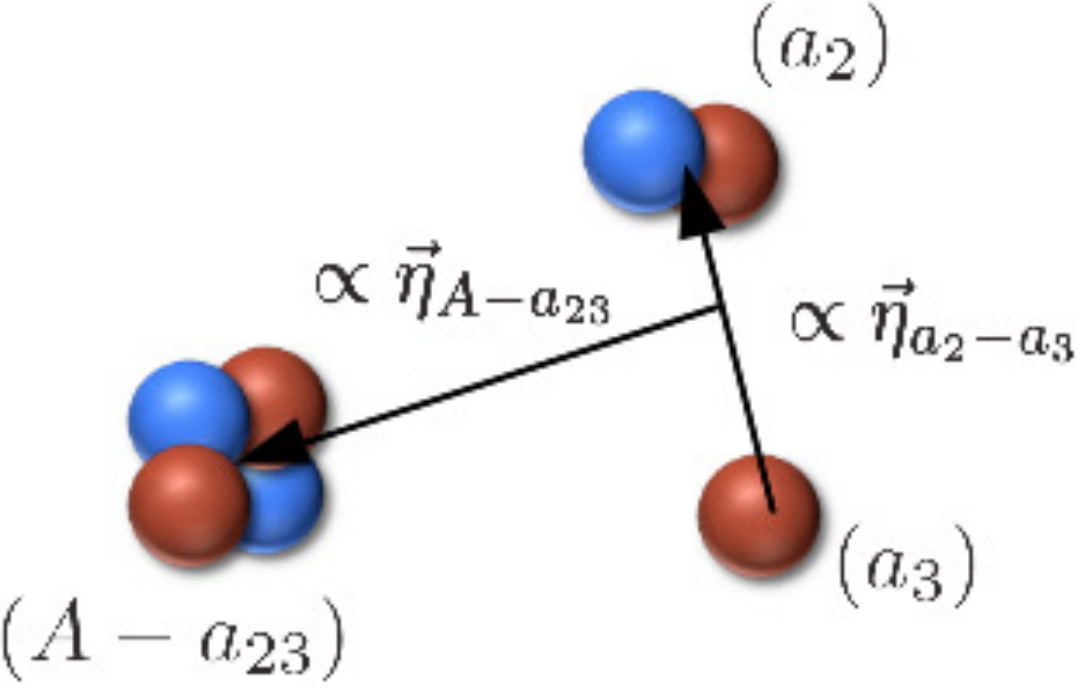}
\caption{(Color online) We show the Jacobi coordinates $\vec\eta_{A-a_{23}}$ (proportional to the vector 
between the c.m.\ of the first cluster and that of the
residual two fragments) and $\vec\eta_{a_2-a_3}$ (proportional to the vector between the c.m.\ of 
clusters 2 and 3). In the
figure, a case with three clusters of four, two and one nucleons are shown, however the formalism is completely
general and can be used to describe any three cluster configuration.}
\label{FigCoor}
\end{figure}

The intrinsic motion of a system of $A$ nucleons arranged into three clusters respectively of mass number $A-a_{23}$, $a_2$, and $a_3$ ($a_{23}=a_2+a_3< A$), can be described by the many-body wave function
\begin{align}
	|\Psi^{J^\pi T}\rangle & = \sum_{\nu} \iint dx \, dy \, x^2\, y^2 \, G_{\nu}^{J^\pi T}(x,y) \, \hat {\mathcal A}_\nu\, |\Phi^{J^\pi T}_{\nu x y} \rangle \,,
	\label{eq:trialwf}
\end{align}
where $G_{\nu}^{J^\pi T}(x,y)$ are continuous variational amplitudes of the integration variables $x$ and $y$, $\hat {\mathcal A}_\nu$ is an appropriate intercluster antisymmetrizer introduced to guarantee the exact preservation of the Pauli exclusion principle, and 
\begin{widetext}
\begin{align}
	 |\Phi^{J^\pi T}_{\nu x y} \rangle  = & 
	\Big[\Big(|A-a_{23}~\alpha_1I_1^{\pi_1}T_1\rangle 
	\left (|a_2\, \alpha_2 I_2^{\pi_2} T_2\rangle |a_3\, \alpha_3 I_3^{\pi_3}T_3\rangle \right)^{(s_{23}T_{23})}\Big)^{(ST)} 
	\left(Y_{\ell_x}(\hat{\eta}_{a_2-a_3})Y_{\ell_y}(\hat{\eta}_{A-a_{23}})\right)^{(L)}\Big]^{(J^{\pi}T)} \nonumber \\
	& \times \frac{\delta(x-\eta_{a_2-a_3})}{x\eta_{a_2-a_3}} \frac{\delta(y-\eta_{A-a_{23}})}{y\eta_{A-a_{23}}}\,,
	\label{eq:3bchannel}	
\end{align}
\end{widetext}
are three-body cluster channels of total angular momentum $J$, parity $\pi$ and isospin $T$.  Here, $|A-a_{23}~\alpha_1I_1^{\pi_1}T_1\rangle$, $|a_2\, \alpha_2 I_2^{\pi_2} T_2\rangle$ and $|a_3\, \alpha_3 I_3^{\pi_3} T_3\rangle$ denote the microscopic (antisymmetric) wave functions of the three nuclear fragments, 
which are labelled by the spin-parity, isospin and energy quantum numbers $I_i^{\pi_i}$, $T_i$, and $\alpha_i$, respectively, with $i=1,2,3$. Additional quantum numbers characterizing the basis states (\ref{eq:3bchannel}) are the spins $\vec s_{23}=\vec I_2 + \vec I_3$ and $\vec S = \vec I_1+ \vec s_{23}$, the orbital angular momenta $\ell_x$, $\ell_y$ and $\vec L = \vec\ell_x+\vec\ell_y$, and the isospin $\vec T_{23}=\vec T_2+\vec T_3$. In our notation, all these quantum numbers are grouped under the cumulative index $\nu = \{A-a_{23}\, \alpha_1I_1^{\pi_1}T_1; a_2\, \alpha_2 I_2^{\pi_2} T_2; a_3\, \alpha_3 I_3^{\pi_3}T_3; s_{23} \,T_{23}\, S \,\ell_x \,\ell_y \, L\}$. Besides the translationally invariant coordinates 
(see e.g.\ Ref.~\cite{PhysRevC.79.044606} Sec.\ II.C) used to describe the internal dynamics of clusters 1, 2 and 3, respectively, in Eq.~(\ref{eq:3bchannel}) we have introduced the Jacobi coordinates $\vec\eta_{A-a_{23}}$ and $\vec\eta_{a_2-a_3}$ where
\begin{align}
	\vec\eta_{A-a_{23}} & = \eta_{A-a_{23}} \hat{\eta}_{A-a_{23}}  \label{eq:etay}\\
	& = \sqrt{\tfrac{a_{23}}{A(A-a_{23})}}  \sum_{i=1}^{A-a_{23}} \vec{r}_i - \sqrt{\tfrac{A-a_{23}}{A\,a_{23}}} \sum_{j=A-a_{23}+1}^A \vec{r}_j \nonumber
\end{align}
is the relative vector proportional to the displacement between the center of mass (c.m.) of the first cluster and that of the residual two fragments, and 
\begin{align}
	\vec\eta_{a_2-a_3} & = \eta_{a_2-a_3} \hat{\eta}_{a_2-a_3} \label{eq:etax}\\
	& =\sqrt{\tfrac{a_3}{a_{23}\,a_2}}  \sum_{i=A-a_{23}+1}^{A-a_3} \vec{r}_i - \sqrt{\tfrac{a_2}{a_{23}\,a_3}} \sum_{j=A-a_3+1}^A \vec{r}_j	\nonumber
\end{align}
is the relative coordinate proportional to the distance between the centers of mass of cluster 
2 and 3 (See figure \ref{FigCoor}). Here, $\vec{r}_i$ denotes the position vector of the $i$-th nucleon.   

Using the expansion~(\ref{eq:trialwf}) for the wave function and projecting the microscopic $A$-body Schr\"odinger equation onto the basis states $\hat {\mathcal A}_\nu\, |\Phi^{J^\pi T}_{\nu x y} \rangle$, the many-body problem can be mapped onto the system of coupled-channel integral-differential equations
\begin{align}
        &\sum_\nu \iint \!\!dx \, dy \, x^2 y^2 \Big [ {\mathcal H}^{J^\pi T}_{\nu^\prime\nu}(x^\prime,y^\prime,x,y) \label{eq:3beq1}\\ 
        &\qquad\qquad\qquad\qquad - E \, {\mathcal N}^{J^\pi T}_{\nu^\prime\nu}(x^\prime,y^\prime,x,y) \Big] G_{\nu}^{J^\pi T}(x,y) = 0\nonumber
\end{align}
for the unknown variational amplitudes $G_{\nu}^{J^\pi T}(x,y)$. Here, $E$ is the total energy of the system in the c.m.\ frame and 
\begin{align}
        {\mathcal H}^{J^\pi T}_{\nu^\prime\nu}(x^\prime,y^\prime,x,y) & = 
                \left\langle\Phi^{J^\pi T}_{\nu^\prime x^\prime y^\prime} \right| \hat {\mathcal A}_{\nu^\prime} H \hat {\mathcal A}_\nu \left | \Phi^{J^\pi T}_{\nu x y} \right\rangle\,, \label{eq:Hkernel} \\
        {\mathcal N}^{J^\pi T}_{\nu^\prime\nu}(x^\prime,y^\prime,x,y) & = 
                \left\langle\Phi^{J^\pi T}_{\nu^\prime x^\prime y^\prime} \right| \hat{\mathcal A}_{\nu^\prime}  \hat {\mathcal A}_\nu \left | \Phi^{J^\pi T}_{\nu x y} \right\rangle   \label{eq:Nkernel}
\end{align}
are integration kernels given respectively by the Hamiltonian and overlap (or norm) matrix elements over the antisymmetrized basis states of Eq.~(\ref{eq:3bchannel}).  Finally, $H$ is the intrinsic $A$-body Hamiltonian. Denoting with $\bar V_C$ the sum of the pairwise average Coulomb interactions among the three clusters in channel $\nu$ of charge numbers $Z_{\nu 1}$, $Z_{\nu 2}$ and $Z_{\nu 3}$, this can be separated into relative-motion and clusters' intrinsic Hamiltonians according to
\begin{align}
        H = T_{\rm rel}+\bar V_C+{\mathcal V}_{\rm rel}+H_{(A-a_{23})}+H_{(a_2)}+H_{(a_3)}\,,
\end{align}
with $T_{\rm rel}$ the relative kinetic energy operator for the three-body system and ${\mathcal V}_{\rm rel}$  the inter-cluster potential given by
\begin{align}
        &{\mathcal V}_{\rm rel} 
        = \sum_{i=1}^{A-a_{23}}\sum_{j=A-a_{23}+1}^A V_{ij} + \sum_{k=A-a_{23}+1}^{A-a_3}\sum_{l=A-a_3+1}^{A} V_{kl} \nonumber\\[3mm]
        &\phantom{{\mathcal V}_{\rm rel} =}+ {\mathcal V}^{3N}_{(A-a_{23},a_2,a_3)} - \bar V_C\,.
        \label{eq:interaction}
\end{align}
Here, ${\mathcal V}^{3N}_{(A-a_{23},a_2,a_3)}$ encompasses the portion of inter-cluster interactions due to the three-nucleon force, which, in general, is part of a realistic Hamiltonian, and $V_{ij}$ is the (nuclear plus point-Coulomb) interaction between nucleons $i$ and $j$. In the present paper we will consider only the nucleon-nucleon ($NN$) component of the inter-cluster interaction and disregard, for the time being, the term  ${\mathcal V}^{3N}_{(A-a_{23},a_2,a_3)}$. The inclusion of the three-nucleon force into the formalism, although computationally much more involved, is straightforward and will be the matter of future investigations. In the remainder of the paper, we will also omit the average Coulomb potential $\bar V_C$, which is null for neutral systems such as the $^4$He+$n$+$n$ investigated here.  The treatment of charged system is nevertheless possible and can be implemented along the same lines of Ref.~\cite{Descouvemont:2005rc}. 
\subsection{Orthogonalized equations}
\label{subsec:orthog}
Owing to the presence of the norm kernel
, the three-cluster equations~(\ref{eq:3beq1}) contain energy-dependent coupling terms. In alternative, one can introduce the orthogonalized Hamiltonian kernel $\bar{\mathcal H}^{J^\pi T}_{\nu^\prime\nu}(x^\prime,y^\prime,x,y)$ of Eq.~(\ref{eq:orthogH})
and solve the more familiar system of multi-channel Schr\"odinger equations
\begin{align}
& \sum_\nu \iint \!\!dx \, dy \, x^2 y^2 \Big [ \bar{\mathcal H}^{J^\pi T}_{\nu^\prime\nu}(x^\prime,y^\prime,x,y) \label{eq:3beq2}\\ 
& \qquad\qquad - E \,\delta_{\nu\nu'}\frac{\delta(x'-x)}{x'x}\frac{\delta(y'-y)}{y'y} \Big] \chi_{\nu}^{J^\pi T}(x,y) = 0\,.
\nonumber
\end{align}   
The amplitudes $G^{J^\pi T}_\nu(x,y)$ of Eq.~(\ref{eq:trialwf}) can then be recovered from the Schr\"odinger wave functions  $\chi^{J^\pi T}_\nu(x,y)$ through Eq.~(\ref{eq:orthogG}).
More details on the orthogonalization procedure can be found in Appendix~\ref{app:ortog}.

\subsection{Hyperspherical Harmonics}
\label{subsec:HH}
The three-cluster Schr\"odinger equations
~(\ref{eq:3beq2}) can be conveniently solved within
the HH basis. This
basis is broadly used \cite{Kievsky:2008es} 
to treat few-body
 problems as its elements are
eigenfunctions of the angular part of the kinetic operator written in 
hyperspherical coordinates \cite{delaripelle83}.  
The first step
is to move to hyperspherical coordinates, i.e. 
\begin{align}
	\eta_{a_2-a_3} = \rho_\eta \sin\alpha_\eta\,, & &x=\rho\sin\alpha \,, \\
	\eta_{A-a_{23}} = \rho_\eta \cos\alpha_\eta\,, & &y=\rho\cos\alpha \,,
\end{align}
where $\rho_\eta$ and $\rho$ are hyperradii and $\alpha_\eta$ and $\alpha$ hyperangles. 
In these coordinates, the relative kinetic energy operator for the three-cluster system can be written as,
\begin{equation}
\hat T_{\rm rel}(\rho)=-\frac{\hbar^2}{2m}\left(\frac{\partial^2}{\partial \rho^2}+\frac{5}{\rho}
\frac{\partial}{\partial \rho}-\frac{\hat{\Lambda}^2(\Omega_\eta)}{\rho^2} \right)
\end{equation}
where the notation $\Omega_\eta$ represents the hyperangle $\alpha_\eta$ and the four angles $\hat\eta_{a_2-a_3}$ and $\hat{\eta}_{A-a_{23}}$ (the direction angles of the Jacobi coordinates $\vec{\eta}_{a_2-a_3}$ and $\vec{\eta}_{A-a_{23}}$, respectively),  $\hat{\Lambda}^2(\Omega_\eta)$ is the Grand-angular kinetic
operator and $m$ is the mass of the nucleon. 

As anticipated, the elements of the HH basis are the eigenfunctions of $\hat{\Lambda}^2(\Omega_\eta)$
\begin{equation}
\mathcal{Y}^{K \ell_x\ell_y}_{L M_L}(\Omega_\eta)=\phi_K^{\ell_x,\ell_y}(\alpha_\eta)\left(
Y_{\ell_x}(\hat\eta_{a_2-a_3})\, Y_{\ell_y}(\hat\eta_{A-a_{23}})\right)^{(L)}_{M_L}
\label{HHbasis}
\end{equation}
with eigenvalues $K(K+4)$. Here, $K$ is the 
hypermomentum quantum number defined as $K=2n+\ell_x+\ell_y$  
with $n=0,1,2,\dotsc$, and the complete set of
functions $\phi_K^{\ell_x,\ell_y}(\alpha)$ is given by
\begin{equation}
\phi_K^{\ell_x,\ell_y}(\alpha)=N_K^{\ell_x\ell_y}(\sin\alpha)^{\ell_x} (\cos \alpha)^{\ell_y}
P_n^{\ell_x+\frac{1}{2},\ell_y+\frac{1}{2}}(\cos 2\alpha)
\label{eq:phi}
\end{equation}
where $P_n^{\alpha,\beta}(\xi)$ are Jacobi polynomials, and $N_K^{\ell_x\ell_y}$ normalization constants.  

As shown in Appendix~\ref{app:HH}, the HH functions~(\ref{HHbasis}) form a natural basis for the description of the three-cluster wave function of Eq.~(\ref{eq:trialwf}) 
and for the solution of the three-cluster dynamical equations. Indeed, by $(i)$ using the expansion
\begin{equation}
\chi_{\nu}^{J^{\pi}T}(\rho,\alpha)=\frac{1}{\rho^{5/2}}\sum_K u^{J^{\pi}T}_{K\nu}(\rho)
\phi_K^{\ell_x,\ell_y}(\alpha)\,
\label{expansion}
\end{equation}
for the relative motion wave functions, where $u^{J^{\pi}T}_{K\nu}(\rho)$  are hyperradial functions 
analogous to those of Eq.~(\ref{eq:chiHH}), and $(ii)$ projecting from the left on the basis states 
$\phi_{K^\prime}^{\ell^\prime_x,\ell^\prime_y}(\alpha^\prime)$, Eq.~(\ref{eq:3beq2}) can be written as a set of non-local 
integral-differential equations in the hyperradial coordinates:
\begin{align}
	\sum_{K\nu}\int d\rho \rho^5 \bar{\cal H}_{\nu'\nu}^{K'K}(\rho',\rho) \frac{u^{J^{\pi}T}_{K\nu}(\rho)}{\rho^{5/2}} 
	= E \frac{u^{J^{\pi}T}_{K^\prime\nu^\prime}(\rho^\prime)}{\rho^{\prime\,5/2}}\,.
	\label{RGMrho}
\end{align}
Here, the orthogonalized Hamiltonian kernel in the hyperradial variables $\rho$ and $\rho^\prime$ is given by
\begin{align}
&\bar{\cal H}_{\nu'\nu}^{K'K}(\rho',\rho)\nonumber \\ 
&\qquad= \int d\alpha' \sin^2\alpha'\cos^2\alpha' \int d\alpha
 \sin^2\alpha\cos^2\alpha  \nonumber \\
&\qquad\phantom{=\int} \times \phi_{K'}^{\ell_x',\ell_y'}(\alpha')\, 
\bar{\mathcal H}^{J^{\pi}T}_{\nu\nu'}(\rho',\alpha',\rho,\alpha) \, \phi_K^{\ell_x,\ell_y}(\alpha)\,.
\label{eq:Hrho}
\end{align}
The solution of Eq.~(\ref{RGMrho}) for the case in which the three clusters form a bound state can be conveniently achieved within the $R$-matrix method as discussed in the next section.

\subsection{Solution of the three-cluster equations for bound states}
\label{subsec:3eq}
We calculate the relative motion wave function by solving Eq.~(\ref{RGMrho})
with the calculable R-matrix method \cite{R-matrix}. 
In particular, we use a Lagrange mesh which simplifies the problem as shown
in many previous works for the two-cluster case
  \cite{PhysRevA.65.052710,BayeJPB98,Hesse2002184,Hesse199837} and has
been generalized to the three-cluster problem in Ref. \cite{Descouvemont:2003ys}.
Within this method, the configuration space is divided into two regions by assuming that the 
Coulomb interaction (if present) is the only interaction experienced by the clusters beyond a finite separation 
$\rho=a$.

In the external region ($\rho>a$), where the Schr\"odinger equation can be solved exactly, the hyperradial wave function is approximated by its known asymptotic form for large $\rho$. For bound states of neutral systems (as the one investigated in this paper), such asymptotic solution is given by:
\begin{equation}
u^{J^{\pi}T}_{K\nu,ext}(\rho)=B_{K\nu}\,\sqrt{k\rho}\,K_{K+2}(k\rho)\,,
\label{external}
\end{equation}
where $K_{K+2}(k\rho)$ is a modified Bessel function of the second kind, $k^2=-2mE/\hbar^2$ is the wave number, and $B_{K\nu}$ is a constant. 
In the internal region ($\rho\le a$), where also the mutual nuclear interaction among the clusters is present,
the wave function is written as a variational expansion on a Lagrange basis of $N$ functions $f_i(\rho)$ 
(see Appendix \ref{lagrange} for definition)
\begin{equation}
u^{J^{\pi}T}_{K\nu,int}(\rho)=\sum_{i=1}^N \beta_{K\nu i} \,f_i(\rho)\,,
\label{internal}
\end{equation}
where $\beta_{K\nu i}$ are the coefficients of the expansion.
The radial wave functions are then obtained by solving in the internal region 
the following set of Bloch-Schr\"odinger equations 
\begin{eqnarray}
\begin{split}
\sum_{K \nu}
\int d\rho \rho^5 
\Bigg(&\bar{\cal H}_{\nu'\nu}^{K'K}(\rho',\rho)
+\mathcal{L}_{K\nu}(\rho)\\ 
&-E\frac{\delta(\rho-\rho')}{\rho^5}\delta_{\nu'\nu}\delta_{K'K}\Bigg)
 \frac{u_{K\nu,int}^{J^{\pi}T}(\rho)}{\rho^{5/2}}\\
&=\sum_{\nu K}
\int d\rho \rho^5 \mathcal{L}_{K\nu}(\rho)\frac{u_{K\nu,ext}^{J^{\pi}T}(\rho)}{\rho^{5/2}}\,,
\end{split}
\label{bloch_eq}
\end{eqnarray}
supplemented by the continuity condition $u_{K\nu,int}^{J^{\pi}T}(a)=u_{K\nu,ext}^{J^{\pi}T}(a)$. Here, we have used the asymptotic expression of Eq.~(\ref{external}) in the right-hand and the  
expansion of Eq.~(\ref{internal}) in the left-hand side of the equation, respectively.  
Further, the elements of the Bloch operator ($L_{K\nu}$ being arbitrary constants)~\cite{Descouvemont:2005rc}
\begin{equation}
\mathcal{L}_{K\nu}(\rho)=\frac{\hbar^2}{2m}\delta(\rho-a)\frac{1}{\rho^{5/2}}\left(\frac{\partial}{\partial\rho}-\frac{L_{K\nu}}{\rho}\right)\rho^{5/2}\,
\label{bloch}
\end{equation}
have the dual function of restoring the hermiticity of the Hamiltonian in the internal region and enforcing the continuity of the derivative of 
the wave function at $\rho=a$~\cite{R-matrix}.
Owing to the Dirac's delta in the Bloch operator, the system of non-local equations~(\ref{bloch_eq}) is equivalent
to that of Eq.~(\ref{RGMrho}) in the internal region. 
Projecting Eq.~(\ref{bloch_eq}) over a basis element $f_{i'}(\rho')$ and choosing the logarithmic derivative 
evaluated in $a$
\begin{equation}
L_{K\nu}(E)=a\frac{u'^{J^{\pi}T}_{K\nu,ext}(a)}{u^{J^{\pi}T}_{K\nu,ext}(a)}
\label{constants}
\end{equation}
for the constants appearing in the definition of the Bloch operator~(\ref{bloch}), 
the system (\ref{bloch_eq}) reduces to
\begin{equation}
\sum_{K'\nu' i'}\left[C^{J^{\pi}T}_{K\nu i,K'\nu' i'}-E\delta_{\nu'\nu}\delta_{K'K}
\delta_{i'i}\right]\beta_{K'\nu' i'}=0\,,
\label{eigen}
\end{equation}
where the elements of the matrix $C^{J^{\pi}T}$ are given by the integrals over the internal region 
\begin{eqnarray}
&&C^{J^{\pi}T}_{K\nu i,K'\nu' i'}=\\ \nonumber
&&\int_0^a d\rho'\int_0^a d\rho \rho^5 
f_{i'}(\rho')\left(\bar{\cal H}_{\nu'\nu}^{K'K}(\rho',\rho)
+\mathcal{L}_{K\nu}\right)f_i(\rho)\,.
\label{eq_C}
\end{eqnarray}

The choice of Lagrange functions as square-integrable basis states for the expansion of the wave 
function in the internal region~(\ref{internal}) greatly simplifies the evaluation of these 
integrals. Indeed, within the Gauss quadrature approximation, the Lagrange functions are 
orthogonal to each other (see Appendix \ref{lagrange}), the matrix elements of nonlocal 
potentials are proportional to the 
values of the nonlocal potentials at the mesh points, and the analytical expression for the 
matrix elements of the kinetic energy operator is straightforward to obtain.  

Note that the matrix $C^{J^{\pi}T}$ depends on the energy, owing to the choice~(\ref{constants}) 
for the boundary conditions in the Bloch operator, which are functions of the wave number $\kappa$ [see Eq.~(\ref{external})]. In practice, the solution of Eq.~(\ref{eq_C}) 
is obtained recursively. One can start from $L_{K\nu}=0$ and iterate the solution of the 
eigenvalue equation~(\ref{eigen}) until the convergence in $E$ is reached, which typically occurs 
in a few iterations. The coefficients of the expansion (\ref{internal}), $\beta_{K\nu i}$, are 
then obtained from the corresponding eigenvector and the relative motion wave functions can be 
constructed using Eqs.~(\ref{expansion}) and
(\ref{internal}).

\subsection{Integration kernels}
\label{subsec:RGMkernels}
The norm  and Hamiltonian integration kernels presented in Sec.~\ref{subsec:microscopic} are calculated within the NCSM/RGM approach as follows.
First, 
the clusters' eigenstates appearing in Eq.~(\ref{eq:3bchannel}) are obtained by diagonalizing the $H_{(A-a_{23})}$, $H_{(a_2)}$ and $H_{(a_3)}$  intrinsic Hamiltonians within the model spaces spanned by the $(A-a_{23})$-, $a_2$- and $a_3$-nucleon NCSM bases, respectively. These are complete sets of many-body HO basis states, the size of which is defined by the maximum number $N_{\rm max}$ of HO quanta above the lowest configuration shared by the nucleons. The same HO frequency $\hbar\Omega$ is used for all three clusters, and the model-space size $N_{\rm max}$ is identical (differs by one) for states of the same (opposite) parity. 

Second, 
for those components that are localized, 
 the matrix elements of the translational invariant
operators $\hat{\mathcal A}_{\nu^\prime}\hat{\mathcal A}_\nu,\,\hat{\mathcal A}_{\nu^\prime}H\hat{\mathcal A}_\nu$  
entering the expression of the integration kernels are evaluated within an HO model space 
using the expansion
\begin{align}
	|\Phi^{J^\pi T}_{\nu x y} \rangle  = 
	&\sum_{n_x n_y}\sum_{Z J_{23}} \hat Z \hat J_{23}\hat S \hat L \,(-1)^{I_1+J_{23}+J+S+Z+\ell_x+\ell_y} \nonumber\\
	&\times 
	\left\{
		\begin{array}{ccc}
			I_1 & s_{23} & S\\[2mm]
			\ell_x & Z & J_{23}\\[2mm]
		\end{array}
	\right\}
	\left\{
		\begin{array}{ccc}
			S & \ell_x & Z\\[2mm]
			\ell_y & J & L\\[2mm]
		\end{array}
	\right\}\nonumber\\[2mm]
	&\times  R_{n_x\ell_x}(x)R_{n_y\ell_y}(y) \, |\Phi^{J^\pi T}_{\gamma n_x n_y} \rangle\,,\label{eq:transf}
\end{align}
where   
$\hat Z = \sqrt{2Z+1}$, $\hat J_{23} = \sqrt{2J_{23}+1}, \cdots$ etc., and $|\Phi^{J^\pi T}_{\gamma n_x n_y} \rangle$ are the HO channel states defined by
\begin{widetext}
\begin{align}
|\Phi_{\gamma n_x n_y}^{J^{\pi}T}\rangle= &\left[\left(|A-a_{23}~\alpha_1 I_1^{\pi_1}T_1\rangle \left(Y_{\ell_x}(\hat{\eta}_{a_2-a_3}) 
\left(|a_2\alpha_2I_2^{\pi_2}T_2\rangle |a_3\alpha_3I_3^{\pi_3}T_3\rangle\right)^{( s_{23} T_{23} )}\right)^{(J_{23}T_{23})}\right)^{(ZT)} 
Y_{\ell_y}(\hat{\eta}_{A-a_{23}})\right]^{(J^{\pi}T)} \nonumber\\[2mm]
&\times R_{n_x\ell_x}(\eta_{a_2-a_3})R_{n_y\ell_y}(\eta_{A-a_{23}})\,,
\label{eq:3bchannelHO}
\end{align}
\end{widetext}
and  labeled by the channel index $\gamma = \{A-a_{23}\, \alpha_1I_1^{\pi_1}T_1; $ $a_2\, \alpha_2 I_2^{\pi_2} T_2; $$a_3\, \alpha_3 I_3^{\pi_3}T_3; \ell_x \,s_{23}J_{23} \,T_{23}\, Z \,\ell_y\}$. Besides the representation of the Dirac's $\delta$ functions of Eq.~(\ref{eq:3bchannel}) in terms of HO radial wave functions $R_{n_x\ell_x}(x)$ and $R_{n_y\ell_y}(y)$
, the transformation of Eq.~(\ref{eq:transf}) reflects a different coupling scheme of the HO channels~(\ref{eq:3bchannelHO}) with respect to the original basis, with $J_{23}$ the total (orbital plus spin) angular momentum quantum number of the system formed by the second and third clusters and $\vec Z = \vec I_1 + \vec J_{23}$ the new channel spin. While the configuration of Eq.~(\ref{eq:3bchannel}) is dictated by the use of the HH as basis for the solution of the three-cluster problem (see Sec.~\ref{subsec:HH} and Appendix~\ref{app:HH}), the binary-cluster-like coupling scheme of Eq.~(\ref{eq:3bchannelHO}) is more convenient for the derivation of the kernels in the HO basis, as it will become clear in a moment.    The frequency $\hbar\Omega$ and the model-space size ($N_{\rm max}/N_{\rm max}+1$ for even/odd parity states) used to expand the relative motion are the same as those adopted for the calculation of the clusters' eigenstates.  

Finally, although the integration kernels are translational invariant quantities, it is computationally convenient to work within a Slater determinant (SD) channel basis $|\Phi^{J^\pi T}_{\gamma n_x n_y} \rangle_{\rm SD}$ defined as in Eq.~(\ref{eq:3bchannelHO}) but with $\vec R^{(a_{23})}_{\rm c.m.}$ in place of the relative vector $\vec \eta_{A-a_{23}}$ and the eigenstates of the heaviest cluster obtained in the SD basis, i.e. 
\begin{align}
&|A-a_{23}~\alpha_1 I_1^{\pi_1}T_1\rangle_{\rm SD}\\
&\quad =|A-a_{23}~\alpha_1 I_1^{\pi_1}T_1\rangle\, R_{00}(R^{(A-a_{23})}_{\rm c.m.})Y_{00}(\hat R^{(A-a_{23})}_{\rm c.m.})\,,\nonumber
\end{align}
where $\vec R^{(A-a_{23})}_{\rm c.m.}$ and $\vec R^{(a_{23})}_{\rm c.m.}$ are respectively the coordinates proportional to the c.m.\ of the first and last two clusters
\begin{align}
	\vec R^{(A-a_{23})}_{\rm c.m.} &= \frac{1}{\sqrt{A-a_{23}}}\sum_{i=1}^{A-a_{23}}\vec r_i\,,\\
      \vec R^{(a_{23})}_{\rm c.m.} &= \frac{1}{\sqrt{a_{23}}}\sum_{j=A-a_{23}+1}^A \vec r_j\,.
\end{align}  
Indeed, the translational invariant matrix elements can be extracted from those calculated in the SD basis, which contain the spurious motion of the c.m., by inverting the following linear transformation: 
\begin{align}
	&\leftsub{\rm SD}{\left\langle\Phi^{J^\pi T}_{\gamma^\prime n_x^\prime n_y^\prime} \right|}\, \hat {\mathcal O}_{\rm t.i.} \left | \Phi^{J^\pi T}_{\gamma n_x n_y} \right\rangle_{\rm SD} \nonumber\\[2mm]
	&\qquad = \sum_{n^{r\prime}_y \ell^{r\prime}_y,n^r_y\ell^r_y,J_r} 
	                   \left\langle\Phi^{J_r^{\pi_r} T}_{\gamma^\prime_r n_x^\prime n_y^{r\prime}} \right|\, \hat {\mathcal O}_{\rm t.i.} \left | \Phi^{J_r^{\pi_r} T}_{\gamma_r n_x n^r_y} \right\rangle\nonumber\\[2mm]
	&\qquad\phantom{=} \times \sum_{\mathcal {N L}} \hat\ell_y\hat\ell^\prime_y\hat J^2_r \, (-1)^{Z+\ell_y-Z^\prime-\ell^\prime_y}\nonumber\\
	&\qquad\phantom{=} \times
	\left\{
		\begin{array}{ccc}
			Z & \ell^r_y & J_r\\[2mm]
			{\mathcal L} & J & \ell_y\\[2mm]
		\end{array}
	\right\}
	\left\{
		\begin{array}{ccc}
			Z^\prime & \ell^{r\prime}_y & J_r\\[2mm]
			{\mathcal L} & J & \ell^\prime_y\\[2mm]
		\end{array}
	\right\}\nonumber\\[2mm]
	&\qquad\phantom{=} \times \langle n^r_y\ell^r_y \,{\mathcal {N L}}\, \ell_y|00\,n_y\ell_y\,\ell_y\rangle_{\frac{a_{23}}{A-a_{23}}}	  
	\nonumber\\[2mm] &\qquad\phantom{=} \times 
      \langle n^{r\prime}_y\ell^{r\prime}_y \,{\mathcal {N L}}\, \ell^\prime_y|00\,n^\prime_y\ell^\prime_y\,\ell^\prime_y\rangle_{\frac{a^\prime_{23}}{A-a^\prime_{23}}} \,. 
      \label{eq:SDtransf}         
\end{align}
Here, $\gamma_r$ denotes a channel index identical to $\gamma$ except for the replacement of the quantum number $\ell_y$ with $\ell^r_y$ (the same applies for the primed indexes) and $\hat{\mathcal O}_{\rm t.i.}$ is any scalar and parity-conserving translational invariant operator.  Further, the transformation~(\ref{eq:SDtransf}) is diagonal in all quantum numbers but $n_y,\ell_y,n^\prime_y,\ell^\prime_y$, and $J^\pi$.  Although formally not strictly necessary, with the new angular momentum coupling scheme of Eq.~(\ref{eq:3bchannelHO}), the present conversion from SD to translational invariant matrix elements represents a straightforward generalization of the analogous binary-cluster transformation discussed in Sec.II.C.2 of Ref.~\cite{PhysRevC.79.044606}, and the most advantageous choice from a computational point of view.

\subsubsection{The (A-2,1,1) mass partition}
The theoretical framework presented so far is general and can in principle be applied to any three-cluster system. In the following, we discuss the derivation of the integration kernels for the more specialized instance of a target nucleus plus two single nucleons ($a_2, a_3=1$), such as the $^4$He+$n$+$n$ system investigated here. Specifically, we will consider the case of identical $(A-2,1,1)$ mass partitions in both the initial and final states.

In this case, the second and third clusters are point-like nucleons with quantum numbers $I^{\pi_{2(3)}}_{2(3)} T_{2(3)} = \tfrac12^+\, \tfrac12$, and the inter-cluster antisymmetrizer is simply given by the product of the antisymmetrization operators for a $(A-2,2)$ mass partition and that of a two-body system
\begin{align}
        \hat{\mathcal A}_\nu & = \hat{\mathcal A}_{(A-2,2)} \, \tfrac{1}{\sqrt2}(1-\hat P_{A-1\,A})\label{eq:A} \\[2mm]
        & = \sqrt{\tfrac{2}{(A-1)A}}\left[ 1\!-\!\sum_{i=1}^{A-2}(\hat P_{iA-1}\!+\!\hat P_{iA})+\!\!\sum_{i<j=1}^{A-2}\hat P_{iA-1}\hat P_{jA}\right]\nonumber\\
        & \phantom{=}\times \tfrac{1}{\sqrt2}\left(1-\hat P_{A-1\,A}\right).\nonumber
\end{align}
Although other factorizations of this operator are of course possible, with the present choice  the antisymmetrization of nucleons $A{-}1$ and $A$ is trivial and can be included in the definition of the channel basis, i.e.
\begin{align}
        \left | \tilde\Phi^{J^\pi T}_{\nu x y} \right\rangle = \frac{ 1-(-1)^{\ell_x+s_{23}+T_{23}}}{\sqrt 2}\left | \Phi^{J^\pi T}_{\nu x y} \right\rangle\,.
        \label{eq:antisym-basis}
\end{align}

The integration kernels for the $(A-2,1,1)$ mass partition are then obtained by evaluating the matrix elements of the operators $ \hat{\mathcal A}^2_{(A-2,2)} =  \sqrt{(A-1)A/2} \,\hat{\mathcal A}_{(A-2,2)}$ and $\hat{\mathcal A}_{(A-2,2)} H \hat{\mathcal A}_{(A-2,2)}$ $ =  \tfrac{1}{2} (\hat{\mathcal A}^2_{(A-2,2)} H + H \hat{\mathcal A}^2_{(A-2,2)})$ on the basis~(\ref{eq:antisym-basis}). For the norm kernel  of Eq.~(\ref{eq:Nkernel}), this yields the following sum of a direct and an exchange term
\begin{align}
        &{\mathcal N}^{J^\pi T}_{\nu^\prime\nu}(x^\prime,y^\prime,x,y)\nonumber\\[2mm]
        &\;\;= \left( 1-(-1)^{\ell_x+s_{23}+T_{23}}\right)\delta_{\nu^\prime\nu}\frac{\delta(x^\prime-x)}{x^\prime x}\frac{\delta(y^\prime-y)}{y^\prime y} \nonumber\\[2mm]
        &\;\;\phantom{=}
        + \, {\mathcal N}^{\rm ex}_{\nu^\prime\nu}(x^\prime,y^\prime,x,y) \,.\label{eq:NAm211}
 \end{align}
Here, the direct term arising from the identical permutation in the antisymmetrization operator 
is calculated in the full space, whereas the non-local exchange term is evaluated within the HO model space. As explained in the previous section, this is achieved by using the expansion~(\ref{eq:transf}), with the translational invariant matrix elements on the HO channel basis of Eq.~(\ref{eq:3bchannelHO}) (antisymmetrized for the exchange of nucleons $A{-}1$ and $A$)
\begin{align}
        &{\mathcal N}^{\rm ex}_{\gamma^\prime n^\prime_x n^\prime_y,\gamma n_x n_y} \label{eq:NexAm211} \\
        &=
        - {2(A-2)} \left\langle\tilde\Phi^{J^\pi T}_{\gamma^\prime n_x^\prime n_y^\prime} \right| \hat P_{A-2\,A} \left | \tilde\Phi^{J^\pi T}_{\gamma n_x n_y} \right\rangle \nonumber\\[2mm]
        &\phantom{=} +\frac{(A-2)(A-3)}{2} \left\langle\tilde\Phi^{J^\pi T}_{\gamma^\prime n_x^\prime n_y^\prime} \right| \hat P_{A-2\,A} \hat P_{A-3\,A-1}\left | \tilde\Phi^{J^ \pi T}_{\gamma n_x n_y} \right\rangle\!. \nonumber
\end{align}
obtained from the corresponding SD ones by inverting
Eq.~(\ref{eq:SDtransf}). At the same time, the calculation of the matrix elements over the SD channels $| \tilde\Phi^{J^\pi T}_{\gamma n_x n_y}\rangle_{\rm SD}$ of Eq.~(\ref{eq:3bchannelHO}) is achieved by first performing a transformation to a fully single-particle basis, i.e.
\begin{align}
        \left| \tilde\Phi^{J^\pi T}_{\gamma n_x n_y}\right\rangle_{\rm SD} & = \!\sum_{a b I L} \!\hat Z \hat I \hat J_{23} \hat s_{23} \hat j_a \hat j_b \hat L^2\,(-1)^{I_1+J+\ell_x+\ell_y+T_{23}} \nonumber\\
        & \phantom{=} \times
        \langle n_a \ell_a\,n_b\ell_b\,L | n_y\ell_y\,n_x\ell_x\,L\rangle_{d=1}\nonumber\\[2mm]
        & \phantom{=} \times
        \left\{
                \begin{array}{ccc}
                        I_1 & J_{23} & Z\\[2mm]
                        \ell_y & J & I\\[2mm]
                \end{array}
        \right\}
        \left\{
                \begin{array}{ccc}
                        \ell_y & L & \ell_x\\[2mm]
                        s_{23} & J_{23} & I\\[2mm]
                \end{array}
        \right\}\nonumber\\
        & \phantom{=} \times
        \left\{
                \begin{array}{ccc}
                        \ell_a & \ell_b &L \\[2mm]
                        \tfrac12 & \tfrac12 & s_{23} \\[2mm]
                        j_a & j_b & I   
                \end{array}
        \right\} \, \left| \Phi^{J^\pi T}_{\kappa_{ab}}\right\rangle_{\rm SD}   \,.
\end{align}
Here, $a$ and $b$ stand for the collections of HO single-particle quantum numbers $\{n_a\ell_aj_a\}$ and $\{n_b\ell_bj_b\}$, respectively, $\langle n_a \ell_a\,n_b\ell_b\,L | n_y\ell_y\,n_x\ell_x\,L\rangle_{d=1}$ indicates an HO bracket for two particles of equal mass, and $\kappa_{ab}=\{A-2 \alpha_1 I^{\pi_1}_1 T_1; $ $n_a\ell_aj_a\tfrac12;n_b\ell_bj_b\tfrac12;IT_{23}\}$ is the index labeling the new SD channel basis
\begin{align}
        |\Phi^{J^\pi T}_{\kappa_{ab}}\rangle_{\rm SD} &= \Big [\left|A-2\, \alpha_1 I_1 T_1\right\rangle_{\rm SD} 
        \left(\varphi_{n_a \ell_a j_a \frac12} (\vec{r}_A \sigma_A \tau_A) \right.
        \nonumber \\
        &\quad\phantom{=}\times \left. \varphi_{n_b \ell_b j_b \frac12} (\vec{r}_{A-1} \sigma_{A-1} \tau_{A-1})\right)^{(I T_{23})}\Big ]^{(J^\pi T)}.
        \label{SD-basis-ab}
\end{align}
We note that,
except for a difference in the notation used for the total isospin of nucleons $A-1$ and $A$, this basis is identical to that introduced for the treatment of binary-cluster channels with a di-nucleon projectile in Eq.~(18) of Ref.~\cite{PhysRevC.83.044609}, where the interested reader can also find the algebraic expressions of the matrix elements
$\leftsub{\rm SD}{\left\langle\Phi^{J^\pi T}_{\kappa_{ab}^\prime}\left| \hat{P}_{A-2\,A} \right|\Phi^{J^\pi T}_{\kappa_{ab}}\right\rangle}_{\rm SD}$
and
$\leftsub{\rm SD}{\left\langle\Phi^{J^\pi T}_{\kappa_{ab}^\prime}\left| \hat{P}_{A-2\,A}\hat{P}_{A-3\,A-1} \right|\Phi^{J^\pi T}_{\kappa_{ab}}\right\rangle}_{\rm SD}$
in Eq.~(19) and (20), respectively. 

The calculation of the Hamiltonian kernel of Eq.~(\ref{eq:Hkernel}) is achieved along the same lines. In this case, the kernel can be divided into a term proportional to the norm kernel discussed above, 
plus a term which resembles the expression of the potential kernel for binary-cluster channels with a di-nucleon projectile (see Ref.~\cite{PhysRevC.83.044609}, Sec.\ II.B): 
\begin{align}
        &\left\langle\tilde\Phi^{J^\pi T}_{\nu^\prime x^\prime y^\prime} \right| H \hat{\mathcal A}^2_{(A-2,2)}\left | \tilde\Phi^{J^\pi T}_{\nu x y} \right\rangle\label{eq:HAm211}\\[2mm]
        &\qquad= \left[ \hat T_{\rm rel}(x^\prime,y^\prime) +\hat V(x^\prime) + E^{I^\prime_1T^\prime_1}_{\alpha^\prime}\right]\,{\mathcal N}^{J^\pi T}_{\nu^\prime\nu}(x^\prime,y^\prime,x,y) \nonumber\\[2mm]
        &\qquad\phantom{=} 
                +\left\langle\tilde\Phi^{J^\pi T}_{\nu^\prime x^\prime y^\prime} \right| {\mathcal V}^{(A-2,2)}_{\rm rel}\hat{\mathcal A}^2_{(A-2,2)}\left | \tilde\Phi^{J^\pi T}_{\nu x y} \right\rangle,\nonumber
                \end{align}
with an analogous expression for the matrix elements of the Hermitian conjugate operator $\hat{\mathcal A}^2_{(A-2,2)} H$.
Here, $\hat V(x^\prime)$ is the potential between nucleons $A$ and $A{-}1$, 
$E^{I^\prime_1T^\prime_1}_{\alpha^\prime}$ is the energy of the $(A{-}2)$-nucleon eigenstate in the final channel, and ${\mathcal V}^{(A-2,2)}_{\rm rel}$ is the sum of pairwise interactions corresponding to the first term in the right-hand side of Eq.~(\ref{eq:interaction}).

As for the exchange operators of the norm, the matrix elements of ${\mathcal V}^{(A-2,2)}_{\rm rel}\hat{\mathcal A}^2_{(A-2,2)}$ are calculated within the HO model space. This involves the evaluation on the SD channel basis of Eq.~(\ref{SD-basis-ab}) of five potential terms: $i)$ $V_{A-2\,A-1}(1{-}\hat P_{A-2\,A-1})$, $ii)$ $V_{A-2\,A}\hat P_{A-2\,A-1}$, $iii)$ $V_{A-3\,A}(1{-}\hat P_{A-3\,A}) \hat P_{A-2\,A-1}$, $iv)$ $V_{A-3\,A-1}\hat P_{A-2\,A-1}$, and $v)$ $V_{A\,A-4}(1{-}\hat P_{A-2\,A-1}) \hat P_{A-3\,A}$. Algebraic expressions for these matrix elements can be found in Eqs. (A1-A4) and (24) of Ref.~\cite{PhysRevC.83.044609}.

Different from the deuteron-nucleus formalism of Ref.~\cite{PhysRevC.83.044609}, where this interaction is already taken into account in the calculation of  the (bound) projectile eigenstate, here the Hamiltonian kernel contains the additional contribution coming from the action of the operator $\hat V(x^\prime)$ on the norm kernel.   In the absence of Coulomb interaction between the last two nucleons (which, if present, can be treated separately as explained in Sec.~\ref{subsec:microscopic}), this term is localized in the variables $x^\prime,x$ and can be calculated as 
\begin{align}
        &\hat V(x^\prime) \, {\mathcal N
}^{J^\pi T}_{\nu^\prime\nu}(x^\prime,y^\prime,x,y)\label{eq:Vx}\\[3mm]
        &\qquad= \hat S \hat S^\prime \hat L \hat L^\prime \, (-1)^{\ell_x+S+L-\ell^\prime_x-S^\prime-L^\prime}\nonumber\\
        &\qquad\phantom{=} \times \sum_{J_{23}} \hat J_{23}^2 
        \left\{
        \begin{array}{cccc}
                - & I_1 & S^\prime & s_{23} \\[2mm]
                \ell_y & - & L^\prime & \ell^\prime_x \\[2mm]
                \ell_x & s_{23} & - & J_{23} \\[2mm]
                L & S & J &-
        \end{array}
        \right\}
        \nonumber\\[2mm]
        &\qquad \phantom{=}\times \sum_{L_x}\sum_{n^\prime_x n_x}R_{n^\prime_x\ell^\prime_x}(x^\prime) R_{n_x L_x}(x)\nonumber\\[2mm]
        &\qquad\phantom{=} \times \langle n^\prime_x \ell^\prime_x s_{23} J_{23} T_{23} | V | n_x L_x s_{23} J_{23} T_{23}\rangle\nonumber\\[2mm]
        &\qquad\phantom{=} \times \left(1-(-1)^{\ell_x+s_{23}+T_{23}}\right)\delta_{\tilde\gamma^\prime \gamma}\frac{\delta(y^\prime-y)}{y^\prime y} \nonumber\\[2mm]
        &\qquad\phantom{=}+ \hat V(x^\prime)\, {\mathcal N
}^{{\rm ex}}_{\nu^\prime\nu}(x^\prime,y^\prime,x,y) \nonumber\,.
\end{align}
Here, the expression between curly brackets represents a 12-j symbol of the second kind (see Appendix \ref{app:12j}), $\langle n^\prime_x \ell^\prime_x s_{23} J_{23} T_{23} | V | n_x L_x s_{23} J_{23} T_{23}\rangle$ are two-body matrix elements of the nuclear interaction on the translational-invariant HO basis, and $\tilde\gamma^\prime$ is an index associated with the HO channel states of Eq.~(\ref{eq:3bchannelHO}) and identical to $\gamma^\prime$ except for the replacement of the quantum number $\ell^\prime_x$ with $L_x$. In the present work, the Dirac's delta function in the $y$ variables of  Eq.~(\ref{eq:Vx}) is approximated by an extended-size expansion in HO radial wave functions that goes well beyond the adopted HO model space ($N_{\rm ext}>>N_{\rm max}$). The influence of such an approximation on the calculated binding energy of $^6$He is small and will be discussed in Sec.~\ref{sec:applications}. 

Finally, with the exception of the terms proportional to the exchange part of the norm kernel, the action of the relative kinetic energy operator $\hat T_{\rm rel}(x^\prime,y^\prime)$ and that of the eigenvalues $E^{I^\prime_1T^\prime_1}_{\alpha^\prime}$ are both calculated in the full space.

\section{\label{sec:applications} Applications to $^6$He}
It is well known that 
$^6$He is the lightest Borromean nucleus~\cite{Tanihata:1995yv, PhysRevLett.55.2676}, formed by an $^4$He core and two halo neutrons. 
Owing to its small mass number and the fact that its constituents do not form bound subsystems, it is an ideal first candidate to be studied within the present approach.

The ground state of this nucleus 
has been the subject of many investigations. Some of them are based on a three-body non-microscopic cluster formalism, representing it as a system of three inert particles~\cite{kukulin86,
Zhukov1993151,Descouvemont:2003ys,Descou12,Descouvemont:2005rc}. This type of three-body 
methods can lead to the appropriate asymptotic behavior of the wave function, but do not allow for the exact treatment of the Pauli principle, which plays a fundamental role for light nuclei, 
and make use of effective nucleon-nucleus potentials. 
There have also been 
{\em ab initio} six-body calculations focused on the ground state of $^6$He 
\cite{Bacca:2009yk,Brida:2010ae,PhysRevC.68.034305,Pieper2001,Brodeur:2012zz}. These are based on realistic Hamiltonians and fulfill the Pauli principle exactly. However, not taking explicitly into account the three-body cluster configuration of this nucleus leads to an
incorrect description of the asymptotic properties of the system.
In between these two approaches are microscopic calculations which take into account both the three-cluster configuration of the system and the internal structure of its constituents,
giving better description of the asymptotic behavior of the nuclear wave 
function while also preserving the Pauli principle \cite{Filippov01091996,PhysRevC.63.034607,Korennov2004249,PhysRevC.80.044310}. Nevertheless, so far these type of calculations have 
been based on semi-realistic interactions, often without spin-orbit force, and on a simplified description of the internal structure of the clusters. 

In this work we present for the 
first time an {\em ab initio} calculation which not only uses realistic interactions
but also takes into account the three-body configuration of this nucleus. 
In particular, we apply the formalism presented in Sec.~\ref{sec:formalism} to study the ground state of $^6$He 
within a $^4$He(g.s.)+$n$+$n$ cluster
basis.
As stated in Sec.~\ref{subsec:RGMkernels}, the $^4$He wave function is calculated within the
NCSM formalism. In the present calculations, we describe the $^4$He core only by its g.s.\ wave function, ignoring its excited states. The inclusion of excited states leads to big technical 
difficulties within the NCSM/RGM formalism because it increases notably 
the number of channels and the calculation becomes unbearable for current 
computational resources. However, this is a minor setback which, once 
the method is established as presented in this work, can be 
overcome by coupling the present three-cluster wave functions with NCSM eigenstates of the six-body system within the NCSMC~\cite{PhysRevLett.110.022505,PhysRevC.87.034326} approach.
For the time being, we estimate core polarization effects, by comparing the computed $J^\pi T=0^+1$ $^4$He(g.s.)+$n$+$n$ g.s.\ energy with that obtained from a NCSM diagonalization of the six-body Hamiltonian. In both calculations, we use the same two-body interaction, namely the SRG evolved~\cite{PhysRevC.75.061001,
PhysRevC.77.064003} potential obtained 
from the chiral N$^3$LO $NN$ interaction \cite{N3LO} with the evolution parameter $\Lambda$=1.5 fm$^{-1}$.

\subsection{$^4$He and $^6$He NCSM calculations}
\label{subsec:He_NCSM}

\begin{figure}[t]
\includegraphics[width=95mm]{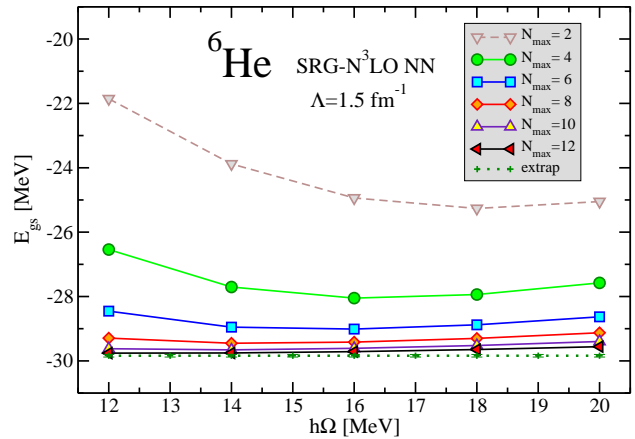}
\caption{(Color online) Convergence pattern of the binding
energy of $^6$He within the NCSM formalism. }
\label{NCSM}
\end{figure}
We performed NCSM calculations for $^4$He that generate eigenstates needed as input for the subsequent three-body cluster NCSM/RGM investigations of $^6$He. Further, we also calculated the g.s.\ energy of $^6$He within the NCSM in order to make a comparison with the $^4$He+$n$+$n$ NCSM/RGM results. The computed $^6$He g.s.\ energies for a range of HO frequencies and various basis sizes ($N_{\rm max}$ values) are presented in Fig.~\ref{NCSM}. As stated earlier, we are employing a soft SRG-evolved chiral N$^3$LO $NN$ interaction with evolution parameter $\Lambda$=1.5 fm$^{-1}$. We intentionally adopt such a soft interaction, for which our calculations reach convergence in the HO basis expansion already at the computationally accessible $N_{\rm max}\sim 12$.  
We can subsequently concentrate on the exploration of the validity of other approximations in the three-cluster NCSM/RGM formalism. We note that the same $NN$ interaction was used in previous binary-cluster NCSM/RGM calculations of the $d$-$^4$He scattering~\cite{PhysRevC.83.044609} and the $d$-$^3$H fusion~\cite{PhysRevLett.108.042503}. The variational NCSM calculations converge rapidly and can be easily extrapolated to $N_{\rm max}\rightarrow\infty$ using, e.g., an exponential function of the type $E(N_{\rm max})=E_\infty+a\; e^{-b N_{\rm max}}$. As shown in Fig.~\ref{NCSM}, at $N_{\rm max}=12$ the dependence of the $^6$He g.s.\ energy on the HO frequency is flat in the range of $\hbar\Omega\sim 12 - 18$ MeV. The variational minimum is close to $\hbar\Omega=14$ MeV that we then choose for the calculation of the $^4$He eigenstates used as input for the $^4$He+$n$+$n$ NCSM/RGM investigations of the $^6$He nucleus.

\begin{table}[b]
\caption{Computed NCSM $^4$He (second column) and NCSM/RGM $^6$He [as $^4$He(g.s.)+$n$+$n$] (third column) g.s.\ energies in MeV as a function of the HO model space size $N_{\rm max}$. The last column shows the $^6$He g.s.\ energies in MeV for model space sizes of $N_{\rm max}-2$. The last two rows show the extrapolated 
values for the calculations with their uncertainties, and the experimental values.}
\begin{ruledtabular}
\begin{tabular}{c c c c}
$N_{\rm max}$ & $^4$He-NCSM &$^6$He-NCSM/RGM & $^6$He-NCSM\\
\hline
6& $-27.984$ & $-28.907$ & $-27.705$ \\
8& $-28.173$ & $-28.616$ & $ -28.952$\\
10& $-28.215$ &  $-28.696$ & $-29.452$\\
12& $-28.224$ & $-28.697$ & $-29.658$ \\
\hline
Extrapolation & $-28.230(5)$ & $-28.70(3)$ & $-29.84(4)$ \\
\hline
Experimental  & $-28.296$& \multicolumn{2}{c}{$-29.268$} \\
\end{tabular}
\end{ruledtabular}
\label{energy}
\end{table}
\begin{figure*}[t]
    \begin{minipage}[c]{7.5cm}\hspace*{-6mm}
      \includegraphics[width=7.5cm,clip=,draft=false]{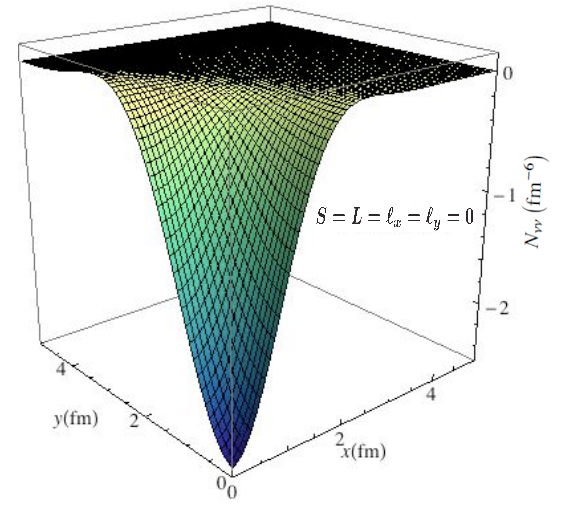}
    \end{minipage}
    \begin{minipage}[c]{7.5cm}\hspace*{-6mm}
      \includegraphics[width=7.5cm,clip=,draft=false]{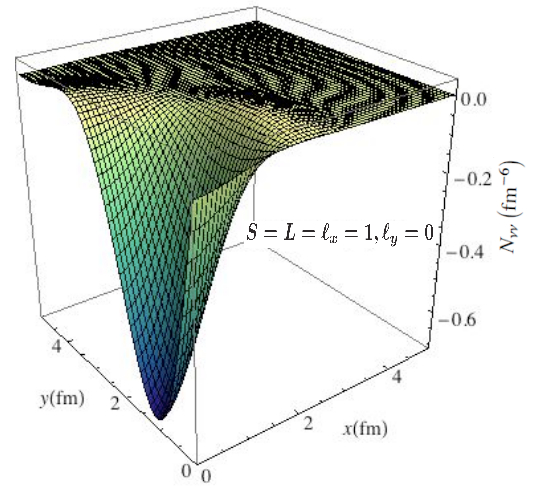}
    \end{minipage}    
    \begin{minipage}[c]{7.5cm}\hspace*{-6mm}
      \includegraphics[width=7.5cm,clip=,draft=false]{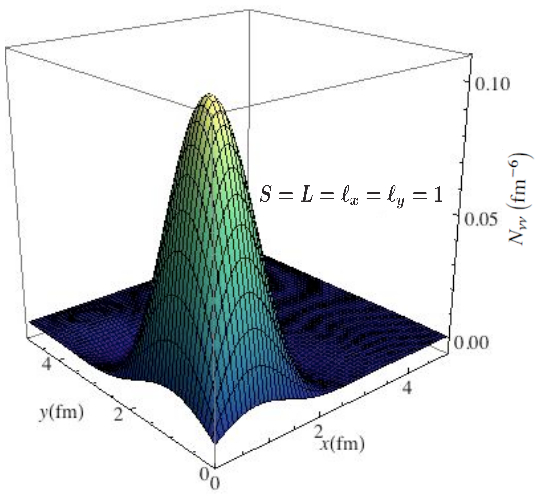}
    \end{minipage}    
    \begin{minipage}[c]{7.5cm}
      \includegraphics[width=7.5cm,clip=,draft=false]{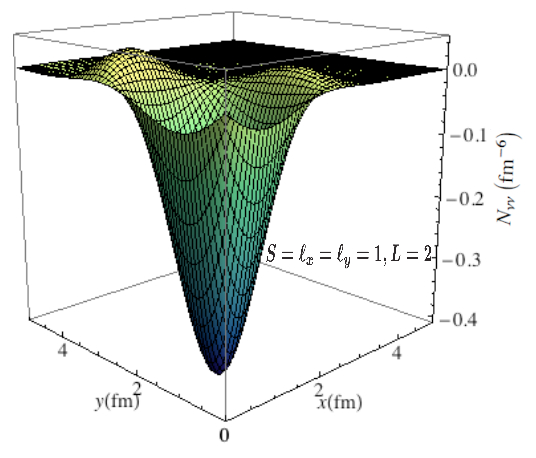}
    \end{minipage}
\caption{(Color online) Diagonal elements of the exchange part of the norm kernel ${\cal N}^{\rm ex}_{\nu\nu}(x,y,x',y')$ for the $S=L=\ell_x=\ell_y=0$ (top left), $S=L=\ell_x=1$, $\ell_y=0$ (top right),  $S=L=\ell_x=\ell_y=1$ (bottom left), and $S=\ell_x=\ell_y=1$, $L=2$ (bottom right) partial waves.  In each plot, the prime $x^\prime$ and $y^\prime$ coordinates are set to 1 fm.\label{norm}}
\end{figure*}
In table \ref{energy}, the energy of
$^4$He and $^6$He ground states calculated with the NCSM formalism 
are shown (in MeV) in the second and last column, respectively. The extrapolated values of the NCSM
calculation with their uncertainties and the experimental values~\cite{Brodeur:2012zz} are also given. It can be observed that, with the present soft SRG interaction,
the g.s.\ energy of $^4$He is close to the experimental value, while the $^6$He ground state is overbound by about 0.5 MeV.

\subsection{$^4$He+$n$+$n$ NCSM/RGM calculations}
We performed calculations for the $^4$He+$n$+$n$ three-cluster system by using the NCSM/RGM formalism described in section \ref{sec:formalism}. In this first application, we neglect core polarization effects and limit the description of $^4$He to just the $I_1^{\pi_1} T_1{=}0^+ 0$ g.s.\  
eigenstate in the NCSM/RGM coupled channel equations. This is the only limitation of the model space introduced. None of the remaining quantum numbers contained in the cumulative index $\nu$ have any restriction other than those dictated by the model space size itself. In particular, we calculated Hamiltonian and norm kernels of Eq.~(\ref{eq:Hkernel}) and (\ref{eq:Nkernel}) for all possible $J^\pi$ channels up to $J=27$, the maximum value of the total angular momentum for our largest model space of $N_{\rm max}=13$, in which both the $\ell_x$ and $\ell_y$ orbital angular momentum quantum numbers can vary from $0$ to $13$. Although, for the present paper we were exclusively interested in the $0^+$ g.s.\ of $^6$He, this was a necessary step to correctly extract the translational invariant matrix elements from our SD calculations through Eq.~(\ref{eq:SDtransf}). Illustrative examples of the norm kernel can be found in Sec.~\ref{subsubsec:norm}.
The g.s.\ energy of $^6$He is then obtained by solving the $^4$He+$n$+$n$ non-local hyperradial equations~(\ref{RGMrho}) for the $J^\pi =0^+$ channel with bound-state boundary conditions, as explained in Sec.~\ref{subsec:3eq}. The dimension of the HH model space used for this part of the calculation is related to the maximum value of the hypermomentum $K_{\rm max}$. Overall, the number of \{$\nu$,$K$\} channels is very large (around 200 for the $J^\pi =0^+$ state alone in our largest model space with $N_{\rm max}=13$ and $K_{\rm max}=28$). The dependence of our results for the $^6$He g.s. with respect to both the HO and HH expansions are discussed in 
Sec.~\ref{subsubsec:He6}.

\subsubsection{Norm kernels}
\label{subsubsec:norm}
Particularly interesting are the elements of the exchange part of the norm kernel, ${\cal N}^{\rm ex}_{\nu'\nu}(x,y,x',y')$, defined in Eq.~(\ref{eq:NAm211}), which give a measure of the influence of the Pauli exclusion principle. In Fig.~\ref{norm}, we present just a few of the most relevant examples. In addition, for visual purposes, we set the value of the primed coordinates $x^\prime$ and $y^\prime$ to $1$ fm. In the present calculation, where the first cluster is given by the g.s.\ of $^4$He, i.e.\ a $I_1^{\pi_1}T_1=0^+0$ state, and the second and third clusters are single nucleons, the various channels can be simply labeled by the spin and orbital angular momentum quantum numbers $S,L,\ell_x$, and $\ell_y$. As one would expect, for the largely $s$-shell $^4$He core the antisymmetrization makes its largest contribution in the $S=L=\ell_x=\ell_y=0$ channel of the $0^+$ state, of which we show the diagonal  ``exchange" norm in the top left panel of Fig.~\ref{norm}. Large negative values of the exchange part of the norm kernel generally correspond to the presence of Pauli forbidden components, in this case the $0\hbar\Omega$ component due to the $s$-wave relative motion in both $x$ and $y$ coordinates. The norm is positive and much smaller in channels where the antisymmetrization plays a minor role, such as the $S=L=\ell_x=\ell_y=1$ displayed in the bottom left panel of the figure. The fairly symmetric appearance of these norm kernels is due to the equal value of the two orbital angular momenta $\ell_x$ and $\ell_y$. However, the kernels are in general asymmetric in $x,y$ ($x^\prime,y^\prime$) as it is particularly evident in the $S=L=\ell_x=1,\ell_y=0$ case (top right panel). In this  component, which appears in the $0^-, 1^-$, and $2^-$ states, one can observe once again the repulsion due to the Pauli principle in the $y$ coordinate, while the $p$-wave motion forces the norm kernel to be null for $x=0$ ($x^\prime=0$). Finally, in the bottom right panel of Fig.~\ref{norm} we present the $S=\ell_x=\ell_y=1, L=2$ diagonal element of the $1^+, 2^+$, and $3^+$ exchange norm. In this channel, where one could naively expect a positive norm, we find a non negligible negative contribution of the antisymmetrization, which suggests the presence of Pauli-forbidden components. These examples show that a correct treatment of the antisymmetrization is not only important to describe the g.s.\ of the $^6$He nucleus, but also plays a role in important excited states such as the $1^-$ or $2^+$ resonances.

\subsubsection{$^6$He ground state}
\label{subsubsec:He6}
\begin{figure}[t]
\includegraphics[width=85mm]{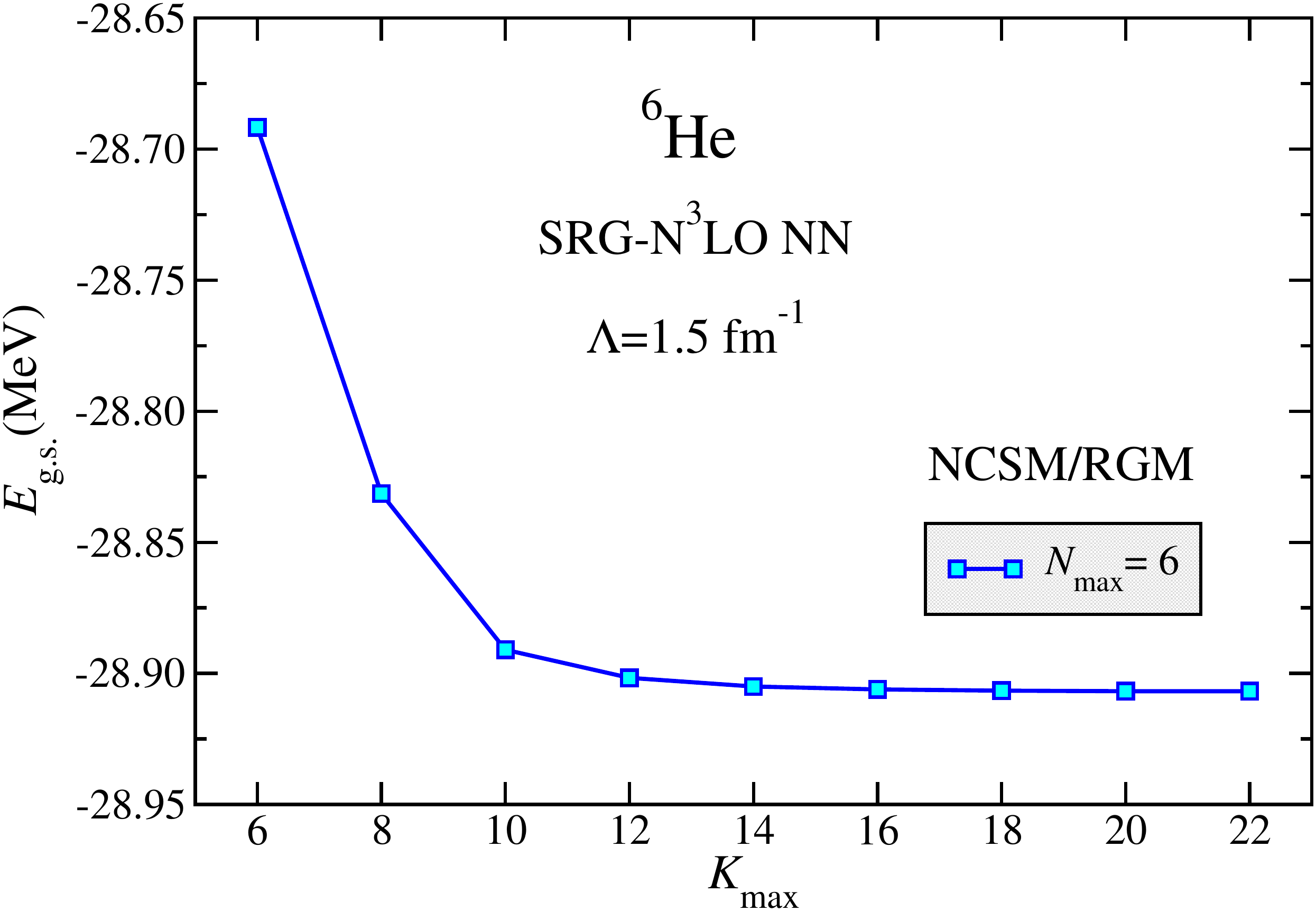}
\caption{(Color online) Dependence of the NCSM/RGM calculated $^6$He g.s.\ energy at $N_{\rm max}=6$ as a function of the maximum value of the hypermomentum $K_{\rm max}$ used in the HH expansion. For these calculations we used a matching radius of $a=20$ fm, $N=30$ Lagrange mesh points, and an extended HO model space of $N_{\rm ext}=40$.}
\label{fig:Kmax}
\end{figure}
\begin{figure}[b]
\includegraphics[width=85mm]{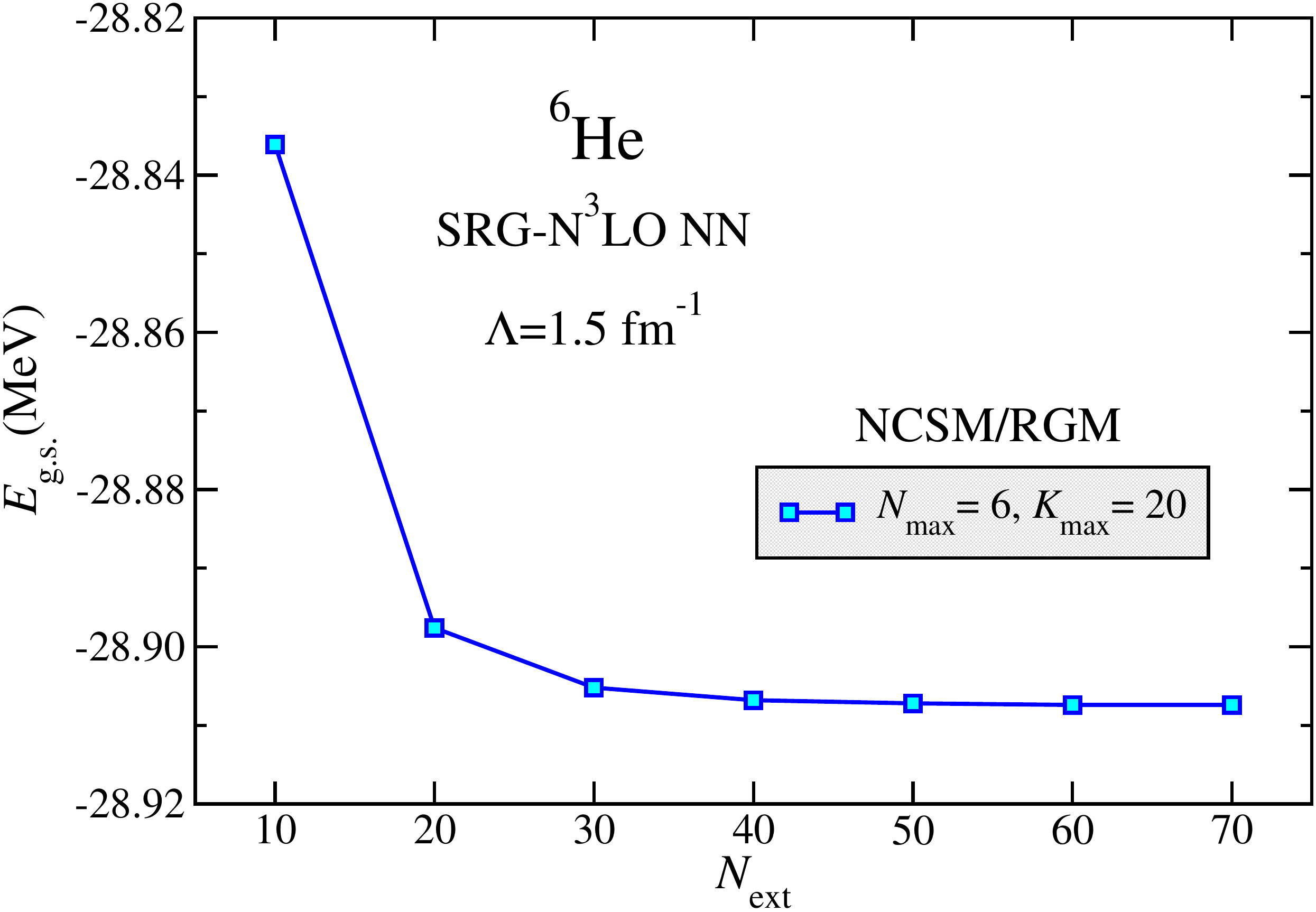}
\caption{(Color online) Dependence of the NCSM/RGM calculated $^6$He g.s.\ energy at $N_{\rm max}=6$  as a function of the size of the extended HO model space $N_{\rm ext}$ used for the calculation of the interaction kernel of Eq.~\ref{eq:Vx}.  For these calculations we used a hyper momentum of $K_{\rm max}=20$, a matching radius of $a=20$ fm, and $N=30$ ($N=40$) Lagrange mesh points for $N_{\rm ext}\le 30$ ($N_{\rm ext}> 30$).}
\label{fig:Next}
\end{figure}
To calculate the g.s.\ of $^6$He, we first orthogonalize the NCSM/RGM equations~(\ref{eq:3beq1}) as explained in Sec.~\ref{subsec:orthog} and  Appendix \ref{app:ortog}. During 
such procedure the eigenvalues
and eigenvectors of the norm kernel in the HO model space are calculated. 
For this $J^\pi T{=}0^+ 1$ state, we observed the appearance of Pauli forbidden norm eigenstates recognizable for their very small eigenvalues 
and negligible overlap with the physical g.s.\ eigenfunction. Spurious states can appear and admix with the low-lying physical eigenstates of the system if such Pauli forbidden norm eigenstates are not eliminated. For the present calculation, we have removed all norm eigenstates with eigenvalues smaller than $0.1$. The unprecedented large number of $\{\nu n\}$ channels ($\sim 300$) is the likely responsible for the occurrence of such unphysical eigenstates of the norm, which had been never observed in our previous binary-cluster NCSM/RGM calculations.
 
We then expanded the orthogonalized NCSM/RGM equation~(\ref{eq:3beq2}) in HH functions and solved the non-local hyperradial equations~(\ref{RGMrho}) for the $^4$He+$n$+$n$ relative motion imposing bound-state boundary conditions, by using the $R$-matrix method on a Lagrange mesh of Sec.~\ref{subsec:3eq}.   
We found a single bound state in the $J^\pi T{=}0^+ 1$ channel and proceeded to study the behavior of our results at fixed $N_{\rm max}$ with respect to the remaining parameters of the calculation. Given the large scale of this computation, we performed this study at $N_{\rm max}=6$. The rate of convergence of the bound state with respect to the size of the adopted HH model space can be judged by examining Fig.~\ref{fig:Kmax}, where we present a study of the 
calculated g.s.\ energy as a function of the maximum value of the hypermomentum $K_{\rm max}$. The results start to stabilize around $K_{\rm max}=14$ and are fully converged already at $K_{\rm max}=20$. At a given $N_{\rm max}$, the calculation is variational in $K_{\rm max}$. Then, we studied the stability of  
the g.s.\ energy with respect to the selection of the matching radius $a$, and we found that it was good as long as we were choosing values larger than 20 fm. The number $N$ of mesh points required for a good convergence of the Lagrange expansion depends on the value of the matching radius. 
For $a = 20$ fm about 30 mesh points are enough, while a larger number is needed if the matching radius is increased. 
\begin{figure}[t]
\includegraphics[width=85mm]{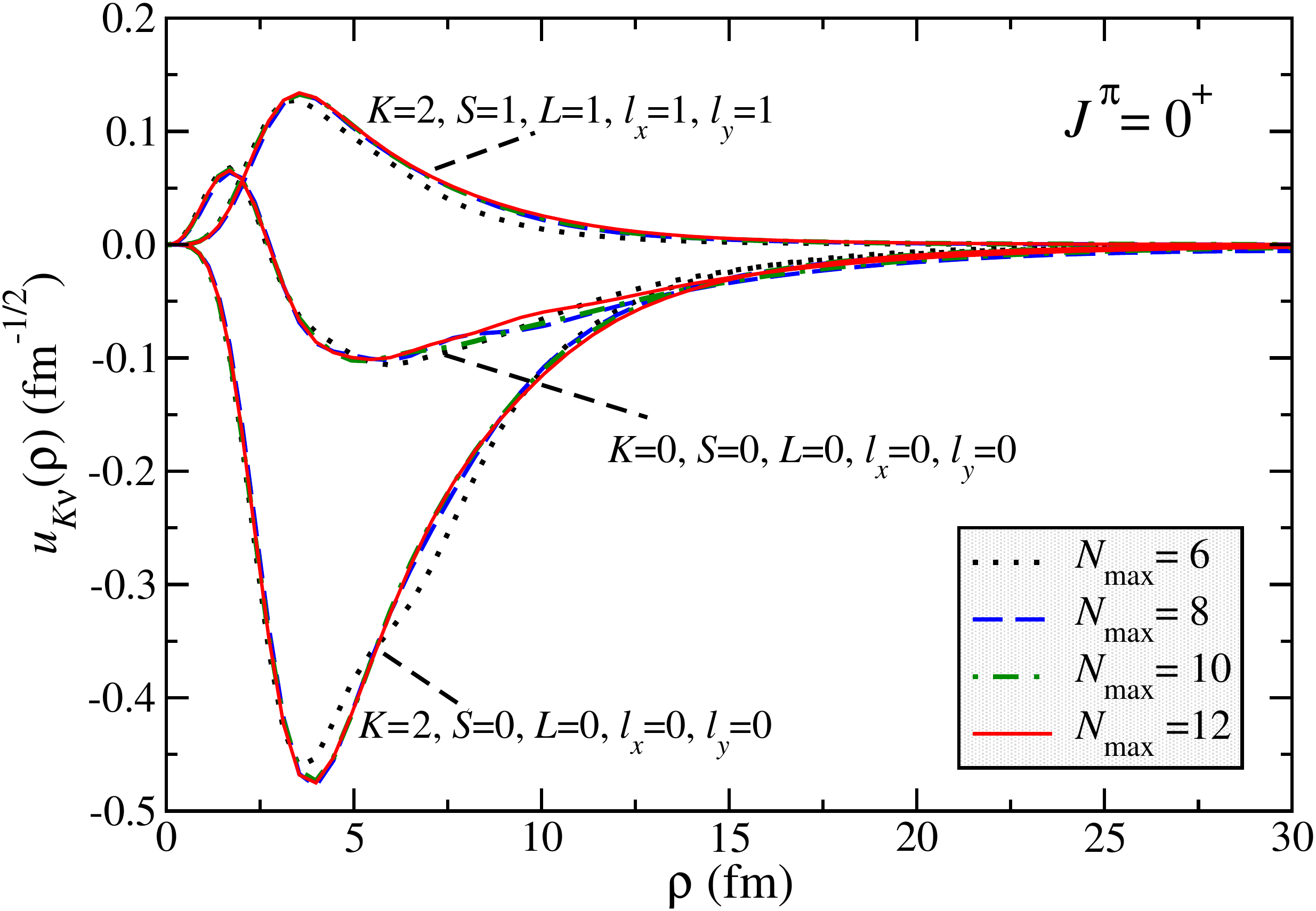}
\caption{(Color online) We show the three main components of the radial part of the $^6$He g.s.\ wave 
functions $u_{K \nu}(\rho)$ for $N_{\rm max}$=6,8,10, and 12. }
\label{f_rho}
\end{figure}
The choice of the $N$ value also depends somewhat on the size of the extended HO model space $N_{\rm ext}$ used to represent the Dirac's delta function 
in the $y$ ($y^\prime$) coordinate (proportional to the distance between the centers of mass of the $^4$He and the two neutrons) while calculating the interaction kernel of Eq.~(\ref{eq:Vx}). Larger $N_{\rm ext}$ values correspond to a larger $y$-range for this potential kernel, which is localized only in the $x$ ($x^\prime$) coordinate. About $30$ ($40$) mesh points are sufficient to reach convergence up to $N_{\rm ext}=30$ ($70$). The behavior of the g.s.\ energy as a function of $N_{\rm ext}$ is presented in Fig.~\ref{fig:Next}. As it can be observed, an extended HO basis size of at least $N_{\rm ext}=40$ is needed to accommodate the long range of this interaction kernel.  Disregarding this effect by computing Eq.~(\ref{eq:Vx}) within the adopted HO model space (i.e. with $N_{\rm ext}=N_{\rm max}$) leads to about $200$ keV underbinding in the $^6$He g.s.\ energy. 
Finally, a stable result for the integrations in the hyperangles $\alpha$ and $\alpha'$ of Eq.~(\ref{eq:Hrho}), which we perform numerically using a Chebyshev-Gauss quadrature (for Chebyshev polynomials of the second kind), was obtained with 20 mesh points. Based on this analysis, we adopted a matching radius of $a=30$ fm with $N=70$ mesh points, a hyper momentum $K_{\rm max}=28$, and an extended HO model space of $N_{\rm ext}=60$ for our larger $N_{\rm max}$ calculations (including the largest with $N_{\rm max}=12)$ presented in the following. 
\begin{figure}[t]
\includegraphics[width=80mm]{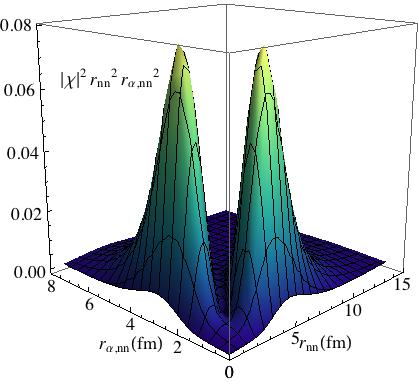}
\caption{(Color online) Probability distribution of the main component of the $^4$He+$n$+$n$
relative motion wave function for the $J^\pi T=0^+1$ ground state. The quantum numbers corresponding to this 
component are $S=L=\ell_x=\ell_y=0$. Here $r_{nn} = \sqrt{2}\, \eta_{nn}$ and $r_{\alpha,nn}=\sqrt{3/4}\,\eta_{\alpha,nn}$ are respectively the distance between the two neutrons and the distance between the c.m. of $^4$He and that of the two neutrons. }
\label{probability}
\end{figure}
In Fig.~\ref{f_rho} the main components of the radial part of the 
relative motion wave function $u_{K\nu}^{J^{\pi}T}$ of the $0^+$ g.s.\
of $^6$He are shown for different 
values of the HO basis size $N_{\rm max}$ used for the expansions of the $^4$He wave function and localized elements of the integration kernels. In the present calculation, each component is uniquely identified by the quantum numbers 
shown in the figure. As it can be seen, 
convergence is almost reached at $N_{\rm max}$=10 and a $N_{\rm max}=14$ calculation, which is currently out of computational reach, is not expected to substantially change the present results. This is confirmed also by the $N_{\rm max}$ dependence of the related g.s.\ energy, presented in the third column of Table~\ref{energy}.  
Contrary to the NCSM, which gives rise to a gaussian asymptotic behavior of the wave function owing to the use of expansions over six-body HO basis states, in the NCSM/RGM the $^4$He(g.s.)+$n$+$n$ wave functions 
present the asymptotic behavior of Eq.~(\ref{external}), which is included by construction
when using the $R$-matrix method. As it can be seen in Fig.~\ref{f_rho}, the 
tails can extend up to about $\rho=25$ fm. This feature will be of great importance when studying $^6$He excited states, using scattering asymptotic conditions in the solution of the three-cluster equations.
\begin{figure}[b]
\includegraphics[width=80mm]{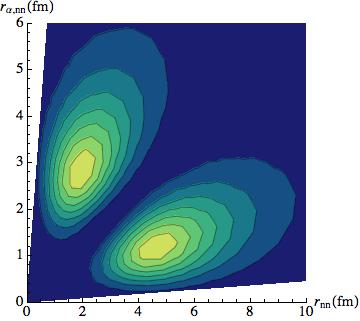}
\caption{(Color online) Contour diagram of the probability distribution 
plotted in Fig. \ref{probability}. }
\label{contour}
\end{figure}

Information about the three-cluster structure of the $^6$He g.s.\ can be obtained by studying the probability distribution 
arising from the main component of the $^4$He+$n$+$n$ relative motion wave function, presented as a surface plot in Fig.~\ref{probability}, and as contour plot in Fig.~\ref{contour}.
This component, characterized by the quantum numbers $S=L=\ell_x=\ell_y=0$, presents the well known~\cite{Descouvemont:2003ys} two-peak 
shape distribution. One peak corresponds to a ``di-neutron" configuration in which the 
neutrons are close together (about 2 fm apart from each other) while the $^4$He core is separated
from their c.m.\ at a distance of about $3$ fm. Whereas the second peak, 
corresponding to the ``cigar" configuration, represents an almost linear
structure in which the two neutrons are far from each other (about $5$ fm apart) and
the alpha particle lies almost in between them at $\sim1$ fm from their center of mass. The
position maxima of the two peaks can be more easily seen from the contour diagram
of 
Fig.~\ref{contour}.
\begin{figure}[b]
\includegraphics[width=80mm]{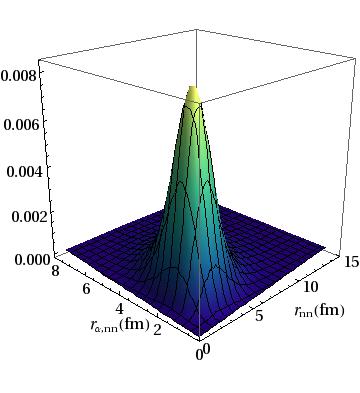}
\caption{(Color online) Same as Fig. \ref{probability}, but for the second 
most important component. The quantum numbers corresponding to this
component are $S=L=\ell_x=\ell_y=1$.}
\label{probability2}
\end{figure}
The second most important contribution to the wave function comes from the 
component with quantum numbers $S=L=\ell_x=\ell_y=1$, and the probability 
distribution arising from it is shown in Fig.~\ref{probability2}. 
From the amplitude of the plot, it can be concluded that this 
component contributes very little to the complete wave function, and does not significantly change the characteristic two peak picture of Fig.~\ref{probability}.
  
The obtained $^6$He g.s.\ energy (in units of MeV) for different sizes of the NCSM/RGM model space $N_{\rm max}$ are presented in the third column of Table~\ref{energy}.
The  results of NCSM calculations of the $^4$He and $^6$He systems are shown
in the second and last column, respectively. 
At $N_{\rm max}=12$, the NCSM/RGM calculation is basically converged within its uncertainty of about $30$ keV, quoted in the last row. The effect of the exclusion of the spurious eigenstates of the norm, discussed at the beginning of this section, 
is included in this uncertainty. Unlike the NCSM case, the present NCSM/RGM calculations are not variational in the HO model space size as at each $N_{\rm max}$ value the three-cluster basis contains a different $^4$He eigenstate. 
It can be observed that in the NCSM/RGM 
the $^6$He ground state is under bound. In particular, the energy is about 1 MeV larger than the one obtained within the NCSM.
This difference is due to excitations of the $^4$He core, which, for technical reasons, are included only in the NCSM calculation at present. In this sense, the  
difference between the two calculations provides a measure of core polarization effects in $^6$He. 

\section{\label{sec:conclusions} Conclusions}

In this work, we have extended the NCSM/RGM method 
\cite{PhysRevLett.101.092501,PhysRevC.79.044606} to  
the treatment of three-cluster dynamics. This new feature permits to study a new range of 
systems that present three-body configurations. In particular, in this work we
show that it can be used to study the structure of
 two-neutron halo nuclei such as $^6$He, and, 
contrary to other {\em ab initio} methods such as the NCSM, to obtain the appropriate asymptotic behavior of its wave functions. Moreover, the present formalism combined with the appropriate scattering boundary conditions
gives access to the {\em ab initio} study of resonant states of two-neutron halo nuclei 
(such as the excited states of $^6$He) as well as to scattering problems involving 
channels with three fragments. Three-cluster NCSM/RGM $^4$He+$n$+$n$ scattering calculations with the aim to study the $^6$He low-lying resonances are currently under way and will be reported in a subsequent paper.

For the present study of $^6$He within the $^4$He+$n$+$n$ cluster basis, we used only the 
ground state wave function to described the $^4$He core. This leads 
to an underbinding of the $^6$He ground state due to the missing treatment of core polarization. 
The difference with respect to the energy obtained from a diagonalization of the Hamiltonian in a NCSM six-body model space
indicates that, with the present soft $NN$ interaction, core-polarization effects amount to less than $5\%$ of the binding. The inclusion of excited states of $^4$He would significantly increase the number of channels in the
calculation, making it computationally unbearable.  Core polarization effects can be more efficiently taken into account by coupling the present three-cluster wave functions with six-body NCSM eigenstates within the NCSMC framework. While the results of this approach will be presented in a forthcoming publication, the main difficulty of such a calculation resides in obtaining the three-cluster NCSM/RGM 
wave functions and was addressed in the present work.

\begin{acknowledgments}
We thank G.\ Hupin many useful discussions. Computing support for this work came from the LLNL institutional Computing Grand Challenge
program and from an INCITE Award on the Titan supercomputer of the Oak Ridge Leadership Computing Facility (OLCF) at ORNL. Prepared in part by LLNL under Contract DE-AC52-07NA27344. Support from the U.S. DOE/SC/NP (Work Proposal No. SCW1158) and
NSERC Grant No. 401945-2011  
is acknowledged. TRIUMF receives funding via a contribution through the 
Canadian National Research Council. 
\end{acknowledgments}

\appendix

\section{Orthogonalization}
\label{app:ortog}

The appearance of the norm kernel ${\cal N}^{J^{\pi}T}_{\nu'\nu}(x',y',x,y)$ in Eq.~(\ref{eq:3beq1}) reflects
the fact that the many-body wave function $\Psi^{J^{\pi}T}$ is expanded in terms of a nonorthogonal basis. 
Therefore, Eq.  (\ref{eq:3beq1}) does not represent a system of multichannel Schr\"odinger equations,
and $G_{\nu}^{J^{\pi}T}(x,y)$ do not represent Schr\"odinger wave functions. 
However, one can solve the equivalent set of orthogonalized equations introduced in Eq.~(\ref{eq:3beq2})
where 
\begin{align}
	\bar{\mathcal H}^{J^\pi T}_{\nu^\prime\nu}(x^\prime,y^\prime,x,y)=\sum_{\tilde\nu'}\iint \!\!  
d\tilde x' d\tilde y' \tilde x'^2   \tilde y'^2 \sum_{\tilde\nu} \iint \!\!
 d\tilde x \,d\tilde y\, \tilde x^2  \tilde y^2  \nonumber \\
\times {\cal N}_{\nu'\tilde\nu'}^{-1/2}(x',y',\tilde x',\tilde y') 
{\mathcal H}^{J^\pi T}_{\tilde\nu'\tilde\nu}(\tilde x',\tilde y',\tilde x,\tilde y)
{\cal N}_{\tilde\nu\nu}^{-1/2}(\tilde x,\tilde y, x, y)\nonumber\\
\label{eq:orthogH}
\end{align}
is the orthogonalized Hamiltonian kernel and the Schr\"odinger wave functions $\chi^{J^\pi T}_\nu(x,y)$ are the new unknowns of the problem,
related to $G^{J^\pi T}_{\nu}(x,y)$ through 
\begin{align}
        &G^{J^\pi T}_\nu(x,y)\label{eq:orthogG} \\
        &\qquad=  \sum_{\tilde\nu}\iint d\tilde x\, d\tilde y \,\tilde x^{2}\,\tilde y^{2} {\mathcal N}^{-\frac12}_{\nu\tilde\nu}(x,y,\tilde x,\tilde y)\,\chi^{J^\pi T}_{\tilde\nu}(\tilde x,\tilde y)\,.\nonumber
\end{align}
Here, ${\cal N}^{1/2}_{\nu'\nu}(x',y',x,y)$ and ${\cal N}^{-1/2}_{\nu'\nu}(x',y',x,y)$ represent the square root and the inverse-square root of the norm kernel, respectively, which are obtained as follows. 
First, we add and subtract from the norm
kernel the identity in the HO model space,
\begin{align}
\label{eq:norm_mod}
&{\cal N}^{J^\pi T}_{\nu'\nu}(x',y',x,y)=\delta_{\nu\nu'}\Bigg[ \frac{\delta(x'-x)}{x'x}\frac{\delta(y'-y)}{y'y}  \nonumber\\
&\qquad\qquad-\sum_{n_xn_y}R_{n_x\ell_x}(x')R_{n_x\ell_x}(x)R_{n_y\ell_y}(y')R_{n_y\ell_y}(y)\Bigg]\nonumber \\[2mm]
&\qquad\qquad+
\Lambda_{\nu^\prime\nu}^{J^{\pi}T}(x^\prime,y^\prime,x,y)\,.
\end{align}
The norm kernel within the truncated 
model space, $\Lambda_{\nu^\prime\nu}^{J^{\pi}T}(x',y',x,y)$, is obtained by using the expansion~(\ref{eq:transf}) with the matrix elements on the HO Jacobi-channel states of Eq. (\ref{eq:3bchannelHO}) given by
\begin{equation}
\Lambda_{\gamma'n_x'n_y',\gamma n_xn_y}^{J^{\pi}T}=\delta_{n_x,n_x'}\delta_{n_y,n_y'}\delta_{\gamma,\gamma'}+
{\mathcal N}^{\rm ex}_{\gamma' n^\prime_x n^\prime_y,\gamma n_x n_y} 
\label{eq:HOnorm}
\end{equation}
and ${\mathcal N}^{\rm ex}_{\gamma' n^\prime_x n^\prime_y,\gamma n_x n_y}$ as defined in Eq.~(\ref{eq:NexAm211}).
Then, the square and the inverse-square root of the full-space norm 
are obtained by $(i)$ finding 
the eigenvalues $\lambda_{\Gamma}$ and eigenvectors $|\varphi_{\Gamma}^{J^{\pi}T}\rangle$ of the matrix 
$\Lambda^{J^{\pi}T}$ of Eq.~(\ref{eq:HOnorm}); $(ii)$ calculating
\begin{equation}
\Lambda_{\gamma'n_x'n_y',\gamma.n_xn_y}^{\pm 1/2}= \sum_{\Gamma} \langle \Phi^{J^{\pi}T}_{\gamma'n_x'n_y'}|
\varphi_{\Gamma}^{J^{\pi}T}\rangle \lambda^{\pm1/2}_{\Gamma}\langle\varphi_{\Gamma}^{J^{\pi}T}|
\Phi^{J^{\pi}T}_{\gamma n_xn_y}\rangle 
\label{eq:lambda_invers}
\end{equation}
and the corresponding integration kernels through the expansion~(\ref{eq:transf}), and finally $(iii)$ replacing the model-space norm $\Lambda^{J^{\pi}T}_{\nu'\nu}(x',y',x,y)$ in Eq.~(\ref{eq:norm_mod}) with $\Lambda_{\nu'\nu}^{\pm 1/2}(x',y',x,y)$, that is,
\begin{align}
&{\cal N}^{\pm 1/2}_{\nu'\nu}(x',y',x,y)=\delta_{\nu\nu'}\Bigg[ \frac{\delta(x'-x)}{x'x}\frac{\delta(y'-y)}{y'y}  \nonumber\\
&\qquad\qquad-\sum_{n_xn_y}R_{n_x\ell_x}(x')R_{n_x\ell_x}(x)R_{n_y\ell_y}(y')R_{n_y\ell_y}(y)\Bigg]\nonumber \\[1mm]
&\qquad\qquad+
\Lambda^{\pm 1/2}_{\nu'\nu}(x',y',x,y)\,. \label{eq:invsqN}
\end{align}
For the inverse operation to be permissible in Eq.~(\ref{eq:lambda_invers}) one has to exclude the subspace 
of (fully) Pauli-forbidden states for which $\lambda_{\Gamma}$=0.
\section{HH channel basis}
\label{app:HH}
Using Dirac delta's properties and completeness relation of the set of $\phi_K^{\ell_x\ell_y}$ functions, 
we have 
\begin{align}
	&\frac{\delta(x-\eta_{a_2-a_3})}{x\,\eta_{a_2-a_3}} \frac{\delta(y-\eta_{A-a_{23}})}{y\,\eta_{A-a_{23}}} \nonumber \\
	&\qquad \quad = \frac{\delta(\rho - \rho_\eta)}{\rho^{5/2}\,\rho_\eta^{5/2}} \,
 \frac{\delta(\alpha - \alpha_\eta)}{\sin\alpha\cos\alpha \, \sin\alpha_\eta\cos\alpha_\eta} \\
 	&\qquad \quad = \frac{\delta(\rho - \rho_\eta)}{\rho^{5/2}\,\rho_\eta^{5/2}} \,
 	\sum_{K}\phi^{\ell_x,\ell_y}_K(\alpha_\eta) \phi^{*\,\ell_x,\ell_y}_K(\alpha),
\end{align}
then the  three-cluster channel states of Eq.~(\ref{eq:3bchannel}) can be written as
\begin{align}
	|\Phi^{J^\pi T}_{\nu x y} \rangle  = \sum_K  \phi^{*\,\ell_x,\ell_y}_K(\alpha) |\Phi^{J^\pi T}_{\nu K \rho} \rangle
	\label{eq:3bchannel2}	
\end{align}
where $|\Phi^{J^\pi T}_{\nu K \rho} \rangle$ are the channel states in the HH basis
\begin{widetext}
\begin{align}
	|\Phi^{J^\pi T}_{\nu K \rho} \rangle = & 
	\Big[\Big(|A-a_{23}~\alpha_1I_1^{\pi_1}T_1\rangle 
	\left (|a_2\, \alpha_2 I_2^{\pi_2} T_2\rangle |a_3\, \alpha_3 I_3^{\pi_3}T_3\rangle \right)^{(s_{23}T_{23})}\Big)^{(ST)}
	{\mathcal Y}_{L}^{K\ell_x\ell_y}(\Omega_\eta)\Big]^{(J^{\pi}T)} \frac{\delta(\rho - \rho_\eta)}{\rho^{5/2}\,\rho_\eta^{5/2}}\,,
	\label{eq:3bchannelHH}	
\end{align}
with $\Omega_\eta = \{\alpha_\eta,\hat\eta_{a_2-a_3},\hat\eta_{A-a_{23}}\}$ and 
${\mathcal Y}_{L}^{K\ell_x\ell_y}(\Omega_\eta)$ the HH basis elements defined in Eq. (\ref{HHbasis}). 
At the same time, inserting this expansion in Eq.~(\ref{eq:trialwf}), 
 changing from $x$ and $y$ to the HH coordinates $\rho$ and $\alpha$ and integrating over the
\end{widetext}
hyperangle $\alpha$, one can demonstrate that the many-body wave function is also given by
\begin{align}
	|\Psi^{J^\pi T}\rangle = \sum_{\nu K} \int d\rho\, \rho^5 \frac{g_{K\nu}^{J^\pi T}(\rho)}{\rho^{5/2}} \hat{\mathcal A}_\nu |\Phi^{J^\pi T}_{\nu K \rho} \rangle\,,
\end{align}
where the hyperradial functions $g_{K\nu}^{J^\pi T}(\rho)$ are obtained from the projection of the variational amplitudes $G^{J^\pi T}_\nu(\rho\sin\alpha,\rho\cos\alpha)$ over the functions $\phi^{\ell_x,\ell_y}_K(\alpha)$:
\begin{align}
	\frac{g_{K\nu}^{J^\pi T}(\rho)}{\rho^{5/2}} = &\int\, d\alpha \sin^2\alpha\cos^2\alpha \nonumber \\
	&\phantom{\int}\times \phi^{*\,\ell_x,\ell_y}_K(\alpha)\, G^{J^\pi T}_\nu(\rho\sin\alpha,\rho\cos\alpha)\,.
	\label{eq:chiHH}
\end{align}

\section{Lagrange basis}
\label{lagrange}

We use a Lagrange basis which is a set of $N$ functions $f_n(x)$ (see \cite{Hesse199837} and references therein), given by 
\begin{equation}
f_n(x)=(-1)^na^{-1/2}\sqrt{\frac{1-x_n}{x_n}}\frac{xP_N(2x/a-1)}{x-ax_n}\,,
\end{equation}
where $P_N(x)$ are Legendre polynomials, and $x_n$ satisfy
\begin{equation}
P_N(2x_n-1)=0\,.
\end{equation}
The Lagrange mesh associated with this basis consists in $N$ points $ax_n$ 
on the interval $[0,a]$ and satisfy the Lagrange conditions
\begin{equation}
f_{n'}(ax_n)=\frac{1}{\sqrt{a\lambda_n}}\delta_{nn'},
\label{LagCond}
\end{equation}
where the coefficients $\lambda_n$ are the weights corresponding to a 
Gauss-Legendre quadrature approximation for the $[0,1]$ interval, i.e.
\begin{equation}
\int_0^1g(x)dx \sim \sum_{n=1}^N \lambda_ng(x_n)\,.
\end{equation}
Using the Lagrange conditions of Eq.~(\ref{LagCond}), it is straightforward to see that 
within the Gauss approximation the Lagrange functions are orthogonal, i.e.
\begin{equation}
\int_0^a f_n(x) f_{n'}(x)dx\sim \delta_{nn'}.
\end{equation}

\begin{widetext}
\section{12-j Symbol definition}
\label{app:12j}
The 12-j symbol of the second kind \cite{varshalovich} is defined by
\begin{align}
       \left\{
        \begin{array}{cccc}
                -   & a_2 & a_3 & a_4 \\[2mm]
                b_1 & -   & b_3 & b_4 \\[2mm]
                c_1 & c_2 & -   & c_4 \\[2mm]
                d_1 & d_2 & d_3 &  -
        \end{array}
        \right\}
&= (-1)^{b_3-a_4-d_1+c_2}\sum_x (2x+1) 
        \left\{
        \begin{array}{ccc}
                a_3 & b_4 & x \\[2mm]
                b_1 & d_3 & b_3 \\[2mm]
        \end{array}
        \right\}
        \left\{
        \begin{array}{ccc}
                a_3 & b_4 & x \\[2mm]
                c_4 & a_2 & a_4 \\[2mm]
        \end{array}
        \right\}  
        \left\{
        \begin{array}{ccc}
                b_1 & d_3 & x \\[2mm]
                d_2 & c_1 & d_1 \\[2mm]
        \end{array}
        \right\}
        \left\{
        \begin{array}{ccc}
                c_4 & a_2 & x \\[2mm]
                d_2 & c_1 & c_2 \\[2mm]
        \end{array}
        \right\} \nonumber \\[4mm]
&= (-1)^{b_3-a_4-d_1+c_2}\sum_x (2x+1) 
        \left\{
        \begin{array}{ccc}
                a_3 & b_3 & d_3 \\[2mm]
                a_4 & b_4 & c_4 \\[2mm]
                a_2 & b_1 & x \\[2mm]
        \end{array}
        \right\}
        \left\{
        \begin{array}{ccc}
                d_2 & d_1 & d_3 \\[2mm]
                c_2 & c_1 & c_4 \\[2mm]
                a_2 & b_1 & x \\[2mm]
        \end{array}
        \right\}\,.  \nonumber \\
\end{align}
\end{widetext}

\bibliographystyle{apsrev4-1}

\begin{thebibliography}{53}%
\makeatletter
\providecommand \@ifxundefined [1]{%
 \@ifx{#1\undefined}
}%
\providecommand \@ifnum [1]{%
 \ifnum #1\expandafter \@firstoftwo
 \else \expandafter \@secondoftwo
 \fi
}%
\providecommand \@ifx [1]{%
 \ifx #1\expandafter \@firstoftwo
 \else \expandafter \@secondoftwo
 \fi
}%
\providecommand \natexlab [1]{#1}%
\providecommand \enquote  [1]{``#1''}%
\providecommand \bibnamefont  [1]{#1}%
\providecommand \bibfnamefont [1]{#1}%
\providecommand \citenamefont [1]{#1}%
\providecommand \href@noop [0]{\@secondoftwo}%
\providecommand \href [0]{\begingroup \@sanitize@url \@href}%
\providecommand \@href[1]{\@@startlink{#1}\@@href}%
\providecommand \@@href[1]{\endgroup#1\@@endlink}%
\providecommand \@sanitize@url [0]{\catcode `\\12\catcode `\$12\catcode
  `\&12\catcode `\#12\catcode `\^12\catcode `\_12\catcode `\%12\relax}%
\providecommand \@@startlink[1]{}%
\providecommand \@@endlink[0]{}%
\providecommand \url  [0]{\begingroup\@sanitize@url \@url }%
\providecommand \@url [1]{\endgroup\@href {#1}{\urlprefix }}%
\providecommand \urlprefix  [0]{URL }%
\providecommand \Eprint [0]{\href }%
\providecommand \doibase [0]{http://dx.doi.org/}%
\providecommand \selectlanguage [0]{\@gobble}%
\providecommand \bibinfo  [0]{\@secondoftwo}%
\providecommand \bibfield  [0]{\@secondoftwo}%
\providecommand \translation [1]{[#1]}%
\providecommand \BibitemOpen [0]{}%
\providecommand \bibitemStop [0]{}%
\providecommand \bibitemNoStop [0]{.\EOS\space}%
\providecommand \EOS [0]{\spacefactor3000\relax}%
\providecommand \BibitemShut  [1]{\csname bibitem#1\endcsname}%
\let\auto@bib@innerbib\@empty
\bibitem [{\citenamefont {Kamada}\ \emph {et~al.}(2001)\citenamefont {Kamada},
  \citenamefont {Nogga}, \citenamefont {Gl\"ockle}, \citenamefont {Hiyama},
  \citenamefont {Kamimura}, \citenamefont {Varga}, \citenamefont {Suzuki},
  \citenamefont {Viviani}, \citenamefont {Kievsky}, \citenamefont {Rosati},
  \citenamefont {Carlson}, \citenamefont {Pieper}, \citenamefont {Wiringa},
  \citenamefont {Navr\'atil}, \citenamefont {Barrett}, \citenamefont {Barnea},
  \citenamefont {Leidemann},\ and\ \citenamefont
  {Orlandini}}]{PhysRevC.64.044001}%
  \BibitemOpen
  \bibfield  {author} {\bibinfo {author} {\bibfnamefont {H.}~\bibnamefont
  {Kamada}}, \bibinfo {author} {\bibfnamefont {A.}~\bibnamefont {Nogga}},
  \bibinfo {author} {\bibfnamefont {W.}~\bibnamefont {Gl\"ockle}}, \bibinfo
  {author} {\bibfnamefont {E.}~\bibnamefont {Hiyama}}, \bibinfo {author}
  {\bibfnamefont {M.}~\bibnamefont {Kamimura}}, \bibinfo {author}
  {\bibfnamefont {K.}~\bibnamefont {Varga}}, \bibinfo {author} {\bibfnamefont
  {Y.}~\bibnamefont {Suzuki}}, \bibinfo {author} {\bibfnamefont
  {M.}~\bibnamefont {Viviani}}, \bibinfo {author} {\bibfnamefont
  {A.}~\bibnamefont {Kievsky}}, \bibinfo {author} {\bibfnamefont
  {S.}~\bibnamefont {Rosati}}, \bibinfo {author} {\bibfnamefont
  {J.}~\bibnamefont {Carlson}}, \bibinfo {author} {\bibfnamefont {S.~C.}\
  \bibnamefont {Pieper}}, \bibinfo {author} {\bibfnamefont {R.~B.}\
  \bibnamefont {Wiringa}}, \bibinfo {author} {\bibfnamefont {P.}~\bibnamefont
  {Navr\'atil}}, \bibinfo {author} {\bibfnamefont {B.~R.}\ \bibnamefont
  {Barrett}}, \bibinfo {author} {\bibfnamefont {N.}~\bibnamefont {Barnea}},
  \bibinfo {author} {\bibfnamefont {W.}~\bibnamefont {Leidemann}}, \ and\
  \bibinfo {author} {\bibfnamefont {G.}~\bibnamefont {Orlandini}},\ }\href
  {\doibase 10.1103/PhysRevC.64.044001} {\bibfield  {journal} {\bibinfo
  {journal} {Phys. Rev. C}\ }\textbf {\bibinfo {volume} {64}},\ \bibinfo
  {pages} {044001} (\bibinfo {year} {2001})}\BibitemShut {NoStop}%
\bibitem [{\citenamefont {Viviani}\ \emph {et~al.}(2011)\citenamefont
  {Viviani}, \citenamefont {Deltuva}, \citenamefont {Lazauskas}, \citenamefont
  {Carbonell}, \citenamefont {Fonseca}, \citenamefont {Kievsky}, \citenamefont
  {Marcucci},\ and\ \citenamefont {Rosati}}]{PhysRevC.84.054010}%
  \BibitemOpen
  \bibfield  {author} {\bibinfo {author} {\bibfnamefont {M.}~\bibnamefont
  {Viviani}}, \bibinfo {author} {\bibfnamefont {A.}~\bibnamefont {Deltuva}},
  \bibinfo {author} {\bibfnamefont {R.}~\bibnamefont {Lazauskas}}, \bibinfo
  {author} {\bibfnamefont {J.}~\bibnamefont {Carbonell}}, \bibinfo {author}
  {\bibfnamefont {A.~C.}\ \bibnamefont {Fonseca}}, \bibinfo {author}
  {\bibfnamefont {A.}~\bibnamefont {Kievsky}}, \bibinfo {author} {\bibfnamefont
  {L.}~\bibnamefont {Marcucci}}, \ and\ \bibinfo {author} {\bibfnamefont
  {S.}~\bibnamefont {Rosati}},\ }\href {\doibase 10.1103/PhysRevC.84.054010}
  {\bibfield  {journal} {\bibinfo  {journal} {Phys. Rev. C}\ }\textbf {\bibinfo
  {volume} {84}},\ \bibinfo {pages} {054010} (\bibinfo {year}
  {2011})}\BibitemShut {NoStop}%
\bibitem [{\citenamefont {Nollett}\ \emph {et~al.}(2007)\citenamefont
  {Nollett}, \citenamefont {Pieper}, \citenamefont {Wiringa}, \citenamefont
  {Carlson},\ and\ \citenamefont {Hale}}]{PhysRevLett.99.022502}%
  \BibitemOpen
  \bibfield  {author} {\bibinfo {author} {\bibfnamefont {K.~M.}\ \bibnamefont
  {Nollett}}, \bibinfo {author} {\bibfnamefont {S.~C.}\ \bibnamefont {Pieper}},
  \bibinfo {author} {\bibfnamefont {R.~B.}\ \bibnamefont {Wiringa}}, \bibinfo
  {author} {\bibfnamefont {J.}~\bibnamefont {Carlson}}, \ and\ \bibinfo
  {author} {\bibfnamefont {G.~M.}\ \bibnamefont {Hale}},\ }\href@noop {}
  {\bibfield  {journal} {\bibinfo  {journal} {Phys. Rev. Lett.}\ }\textbf
  {\bibinfo {volume} {99}},\ \bibinfo {pages} {022502} (\bibinfo {year}
  {2007})}\BibitemShut {NoStop}%
\bibitem [{\citenamefont {Nollett}\ and\ \citenamefont
  {Wiringa}(2011)}]{PhysRevC.83.041001}%
  \BibitemOpen
  \bibfield  {author} {\bibinfo {author} {\bibfnamefont {K.~M.}\ \bibnamefont
  {Nollett}}\ and\ \bibinfo {author} {\bibfnamefont {R.~B.}\ \bibnamefont
  {Wiringa}},\ }\href {\doibase 10.1103/PhysRevC.83.041001} {\bibfield
  {journal} {\bibinfo  {journal} {Phys. Rev. C}\ }\textbf {\bibinfo {volume}
  {83}},\ \bibinfo {pages} {041001} (\bibinfo {year} {2011})}\BibitemShut
  {NoStop}%
\bibitem [{\citenamefont {Nollett}(2012)}]{PhysRevC.86.044330}%
  \BibitemOpen
  \bibfield  {author} {\bibinfo {author} {\bibfnamefont {K.~M.}\ \bibnamefont
  {Nollett}},\ }\href {\doibase 10.1103/PhysRevC.86.044330} {\bibfield
  {journal} {\bibinfo  {journal} {Phys. Rev. C}\ }\textbf {\bibinfo {volume}
  {86}},\ \bibinfo {pages} {044330} (\bibinfo {year} {2012})}\BibitemShut
  {NoStop}%
\bibitem [{\citenamefont {Hagen}\ \emph {et~al.}(2007)\citenamefont {Hagen},
  \citenamefont {Dean}, \citenamefont {Hjorth-Jensen},\ and\ \citenamefont
  {Papenbrock}}]{Hagen2007169}%
  \BibitemOpen
  \bibfield  {author} {\bibinfo {author} {\bibfnamefont {G.}~\bibnamefont
  {Hagen}}, \bibinfo {author} {\bibfnamefont {D.}~\bibnamefont {Dean}},
  \bibinfo {author} {\bibfnamefont {M.}~\bibnamefont {Hjorth-Jensen}}, \ and\
  \bibinfo {author} {\bibfnamefont {T.}~\bibnamefont {Papenbrock}},\ }\href
  {\doibase http://dx.doi.org/10.1016/j.physletb.2007.07.072} {\bibfield
  {journal} {\bibinfo  {journal} {Phys. Lett.}\ }\textbf {\bibinfo {volume}
  {B656}},\ \bibinfo {pages} {169 } (\bibinfo {year} {2007})}\BibitemShut
  {NoStop}%
\bibitem [{\citenamefont {Hagen}\ \emph {et~al.}(2010)\citenamefont {Hagen},
  \citenamefont {Papenbrock},\ and\ \citenamefont
  {Hjorth-Jensen}}]{PhysRevLett.104.182501}%
  \BibitemOpen
  \bibfield  {author} {\bibinfo {author} {\bibfnamefont {G.}~\bibnamefont
  {Hagen}}, \bibinfo {author} {\bibfnamefont {T.}~\bibnamefont {Papenbrock}}, \
  and\ \bibinfo {author} {\bibfnamefont {M.}~\bibnamefont {Hjorth-Jensen}},\
  }\href@noop {} {\bibfield  {journal} {\bibinfo  {journal} {Phys. Rev. Lett.}\
  }\textbf {\bibinfo {volume} {104}},\ \bibinfo {pages} {182501} (\bibinfo
  {year} {2010})}\BibitemShut {NoStop}%
\bibitem [{\citenamefont {Hagen}\ and\ \citenamefont
  {Michel}(2012)}]{PhysRevC.86.021602}%
  \BibitemOpen
  \bibfield  {author} {\bibinfo {author} {\bibfnamefont {G.}~\bibnamefont
  {Hagen}}\ and\ \bibinfo {author} {\bibfnamefont {N.}~\bibnamefont {Michel}},\
  }\href {\doibase 10.1103/PhysRevC.86.021602} {\bibfield  {journal} {\bibinfo
  {journal} {Phys. Rev. C}\ }\textbf {\bibinfo {volume} {86}},\ \bibinfo
  {pages} {021602} (\bibinfo {year} {2012})}\BibitemShut {NoStop}%
\bibitem [{\citenamefont {Baroni}\ \emph
  {et~al.}(2013{\natexlab{a}})\citenamefont {Baroni}, \citenamefont
  {Navr\'atil},\ and\ \citenamefont {Quaglioni}}]{PhysRevLett.110.022505}%
  \BibitemOpen
  \bibfield  {author} {\bibinfo {author} {\bibfnamefont {S.}~\bibnamefont
  {Baroni}}, \bibinfo {author} {\bibfnamefont {P.}~\bibnamefont {Navr\'atil}},
  \ and\ \bibinfo {author} {\bibfnamefont {S.}~\bibnamefont {Quaglioni}},\
  }\href {\doibase 10.1103/PhysRevLett.110.022505} {\bibfield  {journal}
  {\bibinfo  {journal} {Phys. Rev. Lett.}\ }\textbf {\bibinfo {volume} {110}},\
  \bibinfo {pages} {022505} (\bibinfo {year} {2013}{\natexlab{a}})}\BibitemShut
  {NoStop}%
\bibitem [{\citenamefont {Baroni}\ \emph
  {et~al.}(2013{\natexlab{b}})\citenamefont {Baroni}, \citenamefont
  {Navr\'atil},\ and\ \citenamefont {Quaglioni}}]{PhysRevC.87.034326}%
  \BibitemOpen
  \bibfield  {author} {\bibinfo {author} {\bibfnamefont {S.}~\bibnamefont
  {Baroni}}, \bibinfo {author} {\bibfnamefont {P.}~\bibnamefont {Navr\'atil}},
  \ and\ \bibinfo {author} {\bibfnamefont {S.}~\bibnamefont {Quaglioni}},\
  }\href {\doibase 10.1103/PhysRevC.87.034326} {\bibfield  {journal} {\bibinfo
  {journal} {Phys. Rev. C}\ }\textbf {\bibinfo {volume} {87}},\ \bibinfo
  {pages} {034326} (\bibinfo {year} {2013}{\natexlab{b}})}\BibitemShut
  {NoStop}%
\bibitem [{\citenamefont {Wildermuth}\ and\ \citenamefont {Tang}(1977)}]{RGM}%
  \BibitemOpen
  \bibfield  {author} {\bibinfo {author} {\bibfnamefont {K.}~\bibnamefont
  {Wildermuth}}\ and\ \bibinfo {author} {\bibfnamefont {Y.~C.}\ \bibnamefont
  {Tang}},\ }\href@noop {} {\emph {\bibinfo {title} {A unified theory of the
  nucleus}}}\ (\bibinfo  {publisher} {Vieweg},\ \bibinfo {address}
  {Braunschweig},\ \bibinfo {year} {1977})\BibitemShut {NoStop}%
\bibitem [{\citenamefont {Tang}\ \emph {et~al.}(1978)\citenamefont {Tang},
  \citenamefont {LeMere},\ and\ \citenamefont {Thompsom}}]{Tang1978167}%
  \BibitemOpen
  \bibfield  {author} {\bibinfo {author} {\bibfnamefont {Y.}~\bibnamefont
  {Tang}}, \bibinfo {author} {\bibfnamefont {M.}~\bibnamefont {LeMere}}, \ and\
  \bibinfo {author} {\bibfnamefont {D.}~\bibnamefont {Thompsom}},\ }\href@noop
  {} {\bibfield  {journal} {\bibinfo  {journal} {Phys. Rep.}\ }\textbf
  {\bibinfo {volume} {47}},\ \bibinfo {pages} {167} (\bibinfo {year}
  {1978})}\BibitemShut {NoStop}%
\bibitem [{\citenamefont {Fliessbach}\ and\ \citenamefont
  {Walliser}(1982)}]{Fliessbach198284}%
  \BibitemOpen
  \bibfield  {author} {\bibinfo {author} {\bibfnamefont {T.}~\bibnamefont
  {Fliessbach}}\ and\ \bibinfo {author} {\bibfnamefont {H.}~\bibnamefont
  {Walliser}},\ }\href {\doibase
  http://dx.doi.org/10.1016/0375-9474(82)90322-0} {\bibfield  {journal}
  {\bibinfo  {journal} {Nucl. Phys.}\ }\textbf {\bibinfo {volume} {A377}},\
  \bibinfo {pages} {84 } (\bibinfo {year} {1982})}\BibitemShut {NoStop}%
\bibitem [{\citenamefont {Langanke}\ and\ \citenamefont
  {Friedrich}(1986)}]{RGM3}%
  \BibitemOpen
  \bibfield  {author} {\bibinfo {author} {\bibfnamefont {K.}~\bibnamefont
  {Langanke}}\ and\ \bibinfo {author} {\bibfnamefont {H.}~\bibnamefont
  {Friedrich}},\ }\href@noop {} {\emph {\bibinfo {title} {Advances in Nucl.
  Phys.}}}\ (\bibinfo  {publisher} {Plenum},\ \bibinfo {address} {New York},\
  \bibinfo {year} {1986})\BibitemShut {NoStop}%
\bibitem [{\citenamefont {Lovas}\ \emph {et~al.}(1998)\citenamefont {Lovas},
  \citenamefont {Liotta}, \citenamefont {Insolia}, \citenamefont {Varga},\ and\
  \citenamefont {Delion}}]{Lovas1998265}%
  \BibitemOpen
  \bibfield  {author} {\bibinfo {author} {\bibfnamefont {R.}~\bibnamefont
  {Lovas}}, \bibinfo {author} {\bibfnamefont {R.}~\bibnamefont {Liotta}},
  \bibinfo {author} {\bibfnamefont {A.}~\bibnamefont {Insolia}}, \bibinfo
  {author} {\bibfnamefont {K.}~\bibnamefont {Varga}}, \ and\ \bibinfo {author}
  {\bibfnamefont {D.}~\bibnamefont {Delion}},\ }\href {\doibase
  http://dx.doi.org/10.1016/S0370-1573(97)00049-5} {\bibfield  {journal}
  {\bibinfo  {journal} {Phys. Rep.}\ }\textbf {\bibinfo {volume} {294}},\
  \bibinfo {pages} {265 } (\bibinfo {year} {1998})}\BibitemShut {NoStop}%
\bibitem [{\citenamefont {Hofmann}\ and\ \citenamefont
  {Hale}(2008)}]{PhysRevC.77.044002}%
  \BibitemOpen
  \bibfield  {author} {\bibinfo {author} {\bibfnamefont {H.~M.}\ \bibnamefont
  {Hofmann}}\ and\ \bibinfo {author} {\bibfnamefont {G.~M.}\ \bibnamefont
  {Hale}},\ }\href {\doibase 10.1103/PhysRevC.77.044002} {\bibfield  {journal}
  {\bibinfo  {journal} {Phys. Rev. C}\ }\textbf {\bibinfo {volume} {77}},\
  \bibinfo {pages} {044002} (\bibinfo {year} {2008})}\BibitemShut {NoStop}%
\bibitem [{\citenamefont {Navr\'atil}\ \emph
  {et~al.}(2000{\natexlab{a}})\citenamefont {Navr\'atil}, \citenamefont
  {Vary},\ and\ \citenamefont {Barrett}}]{PhysRevLett.84.5728}%
  \BibitemOpen
  \bibfield  {author} {\bibinfo {author} {\bibfnamefont {P.}~\bibnamefont
  {Navr\'atil}}, \bibinfo {author} {\bibfnamefont {J.~P.}\ \bibnamefont
  {Vary}}, \ and\ \bibinfo {author} {\bibfnamefont {B.~R.}\ \bibnamefont
  {Barrett}},\ }\href@noop {} {\bibfield  {journal} {\bibinfo  {journal} {Phys.
  Rev. Lett.}\ }\textbf {\bibinfo {volume} {84}},\ \bibinfo {pages} {5728}
  (\bibinfo {year} {2000}{\natexlab{a}})}\BibitemShut {NoStop}%
\bibitem [{\citenamefont {Navr\'atil}\ \emph
  {et~al.}(2000{\natexlab{b}})\citenamefont {Navr\'atil}, \citenamefont
  {Vary},\ and\ \citenamefont {Barrett}}]{PhysRevC.62.054311}%
  \BibitemOpen
  \bibfield  {author} {\bibinfo {author} {\bibfnamefont {P.}~\bibnamefont
  {Navr\'atil}}, \bibinfo {author} {\bibfnamefont {J.~P.}\ \bibnamefont
  {Vary}}, \ and\ \bibinfo {author} {\bibfnamefont {B.~R.}\ \bibnamefont
  {Barrett}},\ }\href {\doibase 10.1103/PhysRevC.62.054311} {\bibfield
  {journal} {\bibinfo  {journal} {Phys. Rev. C}\ }\textbf {\bibinfo {volume}
  {62}},\ \bibinfo {pages} {054311} (\bibinfo {year}
  {2000}{\natexlab{b}})}\BibitemShut {NoStop}%
\bibitem [{\citenamefont {Quaglioni}\ and\ \citenamefont
  {Navr\'atil}(2008)}]{PhysRevLett.101.092501}%
  \BibitemOpen
  \bibfield  {author} {\bibinfo {author} {\bibfnamefont {S.}~\bibnamefont
  {Quaglioni}}\ and\ \bibinfo {author} {\bibfnamefont {P.}~\bibnamefont
  {Navr\'atil}},\ }\href@noop {} {\bibfield  {journal} {\bibinfo  {journal}
  {Phys. Rev. Lett.}\ }\textbf {\bibinfo {volume} {101}},\ \bibinfo {pages}
  {092501} (\bibinfo {year} {2008})}\BibitemShut {NoStop}%
\bibitem [{\citenamefont {Quaglioni}\ and\ \citenamefont
  {Navr\'atil}(2009)}]{PhysRevC.79.044606}%
  \BibitemOpen
  \bibfield  {author} {\bibinfo {author} {\bibfnamefont {S.}~\bibnamefont
  {Quaglioni}}\ and\ \bibinfo {author} {\bibfnamefont {P.}~\bibnamefont
  {Navr\'atil}},\ }\href@noop {} {\bibfield  {journal} {\bibinfo  {journal}
  {Phys. Rev. C}\ }\textbf {\bibinfo {volume} {79}},\ \bibinfo {pages} {044606}
  (\bibinfo {year} {2009})}\BibitemShut {NoStop}%
\bibitem [{\citenamefont {Navr\'atil}\ \emph {et~al.}(2010)\citenamefont
  {Navr\'atil}, \citenamefont {Roth},\ and\ \citenamefont
  {Quaglioni}}]{PhysRevC.82.034609}%
  \BibitemOpen
  \bibfield  {author} {\bibinfo {author} {\bibfnamefont {P.}~\bibnamefont
  {Navr\'atil}}, \bibinfo {author} {\bibfnamefont {R.}~\bibnamefont {Roth}}, \
  and\ \bibinfo {author} {\bibfnamefont {S.}~\bibnamefont {Quaglioni}},\
  }\href@noop {} {\bibfield  {journal} {\bibinfo  {journal} {Phys. Rev. C}\
  }\textbf {\bibinfo {volume} {82}},\ \bibinfo {pages} {034609} (\bibinfo
  {year} {2010})}\BibitemShut {NoStop}%
\bibitem [{\citenamefont {Navr\'atil}\ and\ \citenamefont
  {Quaglioni}(2011)}]{PhysRevC.83.044609}%
  \BibitemOpen
  \bibfield  {author} {\bibinfo {author} {\bibfnamefont {P.}~\bibnamefont
  {Navr\'atil}}\ and\ \bibinfo {author} {\bibfnamefont {S.}~\bibnamefont
  {Quaglioni}},\ }\href@noop {} {\bibfield  {journal} {\bibinfo  {journal}
  {Phys. Rev. C}\ }\textbf {\bibinfo {volume} {83}},\ \bibinfo {pages} {044609}
  (\bibinfo {year} {2011})}\BibitemShut {NoStop}%
\bibitem [{\citenamefont {Navr{\'a}til}\ \emph {et~al.}(2011)\citenamefont
  {Navr{\'a}til}, \citenamefont {Roth},\ and\ \citenamefont
  {Quaglioni}}]{Navratil2011379}%
  \BibitemOpen
  \bibfield  {author} {\bibinfo {author} {\bibfnamefont {P.}~\bibnamefont
  {Navr{\'a}til}}, \bibinfo {author} {\bibfnamefont {R.}~\bibnamefont {Roth}},
  \ and\ \bibinfo {author} {\bibfnamefont {S.}~\bibnamefont {Quaglioni}},\
  }\href@noop {} {\bibfield  {journal} {\bibinfo  {journal} {Phys. Lett.}\
  }\textbf {\bibinfo {volume} {B704}},\ \bibinfo {pages} {379 } (\bibinfo
  {year} {2011})}\BibitemShut {NoStop}%
\bibitem [{\citenamefont {Navr\'atil}\ and\ \citenamefont
  {Quaglioni}(2012)}]{PhysRevLett.108.042503}%
  \BibitemOpen
  \bibfield  {author} {\bibinfo {author} {\bibfnamefont {P.}~\bibnamefont
  {Navr\'atil}}\ and\ \bibinfo {author} {\bibfnamefont {S.}~\bibnamefont
  {Quaglioni}},\ }\href@noop {} {\bibfield  {journal} {\bibinfo  {journal}
  {Phys. Rev. Lett.}\ }\textbf {\bibinfo {volume} {108}},\ \bibinfo {pages}
  {042503} (\bibinfo {year} {2012})}\BibitemShut {NoStop}%
\bibitem [{\citenamefont {Filippov}\ \emph {et~al.}(1996)\citenamefont
  {Filippov}, \citenamefont {Kato},\ and\ \citenamefont
  {Korennov}}]{Filippov01091996}%
  \BibitemOpen
  \bibfield  {author} {\bibinfo {author} {\bibfnamefont {G.~F.}\ \bibnamefont
  {Filippov}}, \bibinfo {author} {\bibfnamefont {K.}~\bibnamefont {Kato}}, \
  and\ \bibinfo {author} {\bibfnamefont {S.~V.}\ \bibnamefont {Korennov}},\
  }\href {\doibase 10.1143/PTP.96.575} {\bibfield  {journal} {\bibinfo
  {journal} {Prog. Theor. Phys.}\ }\textbf {\bibinfo {volume} {96}},\ \bibinfo
  {pages} {575} (\bibinfo {year} {1996})}\BibitemShut {NoStop}%
\bibitem [{\citenamefont {Vasilevsky}\ \emph
  {et~al.}(2001{\natexlab{a}})\citenamefont {Vasilevsky}, \citenamefont
  {Nesterov}, \citenamefont {Arickx},\ and\ \citenamefont
  {Broeckhove}}]{PhysRevC.63.034606}%
  \BibitemOpen
  \bibfield  {author} {\bibinfo {author} {\bibfnamefont {V.}~\bibnamefont
  {Vasilevsky}}, \bibinfo {author} {\bibfnamefont {A.~V.}\ \bibnamefont
  {Nesterov}}, \bibinfo {author} {\bibfnamefont {F.}~\bibnamefont {Arickx}}, \
  and\ \bibinfo {author} {\bibfnamefont {J.}~\bibnamefont {Broeckhove}},\
  }\href {\doibase 10.1103/PhysRevC.63.034606} {\bibfield  {journal} {\bibinfo
  {journal} {Phys. Rev. C}\ }\textbf {\bibinfo {volume} {63}},\ \bibinfo
  {pages} {034606} (\bibinfo {year} {2001}{\natexlab{a}})}\BibitemShut
  {NoStop}%
\bibitem [{\citenamefont {Vasilevsky}\ \emph
  {et~al.}(2001{\natexlab{b}})\citenamefont {Vasilevsky}, \citenamefont
  {Nesterov}, \citenamefont {Arickx},\ and\ \citenamefont
  {Broeckhove}}]{PhysRevC.63.034607}%
  \BibitemOpen
  \bibfield  {author} {\bibinfo {author} {\bibfnamefont {V.}~\bibnamefont
  {Vasilevsky}}, \bibinfo {author} {\bibfnamefont {A.~V.}\ \bibnamefont
  {Nesterov}}, \bibinfo {author} {\bibfnamefont {F.}~\bibnamefont {Arickx}}, \
  and\ \bibinfo {author} {\bibfnamefont {J.}~\bibnamefont {Broeckhove}},\
  }\href {\doibase 10.1103/PhysRevC.63.034607} {\bibfield  {journal} {\bibinfo
  {journal} {Phys. Rev. C}\ }\textbf {\bibinfo {volume} {63}},\ \bibinfo
  {pages} {034607} (\bibinfo {year} {2001}{\natexlab{b}})}\BibitemShut
  {NoStop}%
\bibitem [{\citenamefont {Korennov}\ and\ \citenamefont
  {Descouvemont}(2004)}]{Korennov2004249}%
  \BibitemOpen
  \bibfield  {author} {\bibinfo {author} {\bibfnamefont {S.}~\bibnamefont
  {Korennov}}\ and\ \bibinfo {author} {\bibfnamefont {P.}~\bibnamefont
  {Descouvemont}},\ }\href {\doibase
  http://dx.doi.org/10.1016/j.nuclphysa.2004.05.013} {\bibfield  {journal}
  {\bibinfo  {journal} {Nucl. Phys.}\ }\textbf {\bibinfo {volume} {A740}},\
  \bibinfo {pages} {249 } (\bibinfo {year} {2004})}\BibitemShut {NoStop}%
\bibitem [{\citenamefont {Damman}\ and\ \citenamefont
  {Descouvemont}(2009)}]{PhysRevC.80.044310}%
  \BibitemOpen
  \bibfield  {author} {\bibinfo {author} {\bibfnamefont {A.}~\bibnamefont
  {Damman}}\ and\ \bibinfo {author} {\bibfnamefont {P.}~\bibnamefont
  {Descouvemont}},\ }\href {\doibase 10.1103/PhysRevC.80.044310} {\bibfield
  {journal} {\bibinfo  {journal} {Phys. Rev. C}\ }\textbf {\bibinfo {volume}
  {80}},\ \bibinfo {pages} {044310} (\bibinfo {year} {2009})}\BibitemShut
  {NoStop}%
\bibitem [{\citenamefont {Vasilevsky}\ \emph {et~al.}(2012)\citenamefont
  {Vasilevsky}, \citenamefont {Arickx}, \citenamefont {Vanroose},\ and\
  \citenamefont {Broeckhove}}]{PhysRevC.85.034318}%
  \BibitemOpen
  \bibfield  {author} {\bibinfo {author} {\bibfnamefont {V.}~\bibnamefont
  {Vasilevsky}}, \bibinfo {author} {\bibfnamefont {F.}~\bibnamefont {Arickx}},
  \bibinfo {author} {\bibfnamefont {W.}~\bibnamefont {Vanroose}}, \ and\
  \bibinfo {author} {\bibfnamefont {J.}~\bibnamefont {Broeckhove}},\ }\href
  {\doibase 10.1103/PhysRevC.85.034318} {\bibfield  {journal} {\bibinfo
  {journal} {Phys. Rev. C}\ }\textbf {\bibinfo {volume} {85}},\ \bibinfo
  {pages} {034318} (\bibinfo {year} {2012})}\BibitemShut {NoStop}%
\bibitem [{\citenamefont {Descouvemont}\ \emph {et~al.}(2003)\citenamefont
  {Descouvemont}, \citenamefont {Daniel},\ and\ \citenamefont
  {Baye}}]{Descouvemont:2003ys}%
  \BibitemOpen
  \bibfield  {author} {\bibinfo {author} {\bibfnamefont {P.}~\bibnamefont
  {Descouvemont}}, \bibinfo {author} {\bibfnamefont {C.}~\bibnamefont
  {Daniel}}, \ and\ \bibinfo {author} {\bibfnamefont {D.}~\bibnamefont
  {Baye}},\ }\href {\doibase 10.1103/PhysRevC.67.044309} {\bibfield  {journal}
  {\bibinfo  {journal} {Phys. Rev. C}\ }\textbf {\bibinfo {volume} {67}},\
  \bibinfo {pages} {044309} (\bibinfo {year} {2003})}\BibitemShut {NoStop}%
\bibitem [{\citenamefont {Descouvemont}\ \emph {et~al.}(2006)\citenamefont
  {Descouvemont}, \citenamefont {Tursunov},\ and\ \citenamefont
  {Baye}}]{Descouvemont:2005rc}%
  \BibitemOpen
  \bibfield  {author} {\bibinfo {author} {\bibfnamefont {P.}~\bibnamefont
  {Descouvemont}}, \bibinfo {author} {\bibfnamefont {E.}~\bibnamefont
  {Tursunov}}, \ and\ \bibinfo {author} {\bibfnamefont {D.}~\bibnamefont
  {Baye}},\ }\href {\doibase 10.1016/j.nuclphysa.2005.11.010} {\bibfield
  {journal} {\bibinfo  {journal} {Nucl. Phys.}\ }\textbf {\bibinfo {volume}
  {A765}},\ \bibinfo {pages} {370} (\bibinfo {year} {2006})}\BibitemShut
  {NoStop}%
\bibitem [{\citenamefont {Bogner}\ \emph {et~al.}(2007)\citenamefont {Bogner},
  \citenamefont {Furnstahl},\ and\ \citenamefont {Perry}}]{PhysRevC.75.061001}%
  \BibitemOpen
  \bibfield  {author} {\bibinfo {author} {\bibfnamefont {S.~K.}\ \bibnamefont
  {Bogner}}, \bibinfo {author} {\bibfnamefont {R.~J.}\ \bibnamefont
  {Furnstahl}}, \ and\ \bibinfo {author} {\bibfnamefont {R.~J.}\ \bibnamefont
  {Perry}},\ }\href@noop {} {\bibfield  {journal} {\bibinfo  {journal} {Phys.
  Rev. C}\ }\textbf {\bibinfo {volume} {75}},\ \bibinfo {pages} {061001}
  (\bibinfo {year} {2007})}\BibitemShut {NoStop}%
\bibitem [{\citenamefont {Roth}\ \emph {et~al.}(2008)\citenamefont {Roth},
  \citenamefont {Reinhardt},\ and\ \citenamefont
  {Hergert}}]{PhysRevC.77.064003}%
  \BibitemOpen
  \bibfield  {author} {\bibinfo {author} {\bibfnamefont {R.}~\bibnamefont
  {Roth}}, \bibinfo {author} {\bibfnamefont {S.}~\bibnamefont {Reinhardt}}, \
  and\ \bibinfo {author} {\bibfnamefont {H.}~\bibnamefont {Hergert}},\
  }\href@noop {} {\bibfield  {journal} {\bibinfo  {journal} {Phys. Rev. C}\
  }\textbf {\bibinfo {volume} {77}},\ \bibinfo {pages} {064003} (\bibinfo
  {year} {2008})}\BibitemShut {NoStop}%
\bibitem [{\citenamefont {Entem}\ and\ \citenamefont {Machleidt}(2003)}]{N3LO}%
  \BibitemOpen
  \bibfield  {author} {\bibinfo {author} {\bibfnamefont {D.~R.}\ \bibnamefont
  {Entem}}\ and\ \bibinfo {author} {\bibfnamefont {R.}~\bibnamefont
  {Machleidt}},\ }\href@noop {} {\bibfield  {journal} {\bibinfo  {journal}
  {Phys. Rev. C}\ }\textbf {\bibinfo {volume} {68}},\ \bibinfo {pages} {041001}
  (\bibinfo {year} {2003})}\BibitemShut {NoStop}%
\bibitem [{\citenamefont {Kievsky}\ \emph {et~al.}(2008)\citenamefont
  {Kievsky}, \citenamefont {Rosati}, \citenamefont {Viviani}, \citenamefont
  {Marcucci},\ and\ \citenamefont {Girlanda}}]{Kievsky:2008es}%
  \BibitemOpen
  \bibfield  {author} {\bibinfo {author} {\bibfnamefont {A.}~\bibnamefont
  {Kievsky}}, \bibinfo {author} {\bibfnamefont {S.}~\bibnamefont {Rosati}},
  \bibinfo {author} {\bibfnamefont {M.}~\bibnamefont {Viviani}}, \bibinfo
  {author} {\bibfnamefont {L.}~\bibnamefont {Marcucci}}, \ and\ \bibinfo
  {author} {\bibfnamefont {L.}~\bibnamefont {Girlanda}},\ }\href {\doibase
  10.1088/0954-3899/35/6/063101} {\bibfield  {journal} {\bibinfo  {journal} {J.
  Phys.}\ }\textbf {\bibinfo {volume} {G35}},\ \bibinfo {pages} {063101}
  (\bibinfo {year} {2008})}\BibitemShut {NoStop}%
\bibitem [{\citenamefont {delaRipelle}(1983)}]{delaripelle83}%
  \BibitemOpen
  \bibfield  {author} {\bibinfo {author} {\bibfnamefont {M.}~\bibnamefont
  {delaRipelle}},\ }\href@noop {} {\bibfield  {journal} {\bibinfo  {journal}
  {Ann. Phys.}\ }\textbf {\bibinfo {volume} {147}},\ \bibinfo {pages} {281}
  (\bibinfo {year} {1983})}\BibitemShut {NoStop}%
\bibitem [{\citenamefont {Descouvemont}\ and\ \citenamefont
  {Baye}(2010)}]{R-matrix}%
  \BibitemOpen
  \bibfield  {author} {\bibinfo {author} {\bibfnamefont {P.}~\bibnamefont
  {Descouvemont}}\ and\ \bibinfo {author} {\bibfnamefont {D.}~\bibnamefont
  {Baye}},\ }\href@noop {} {\bibfield  {journal} {\bibinfo  {journal} {Rep.
  Prog. Phys.}\ }\textbf {\bibinfo {volume} {73}},\ \bibinfo {pages} {036301}
  (\bibinfo {year} {2010})}\BibitemShut {NoStop}%
\bibitem [{\citenamefont {Baye}\ \emph {et~al.}(2002)\citenamefont {Baye},
  \citenamefont {Goldbeter},\ and\ \citenamefont
  {Sparenberg}}]{PhysRevA.65.052710}%
  \BibitemOpen
  \bibfield  {author} {\bibinfo {author} {\bibfnamefont {D.}~\bibnamefont
  {Baye}}, \bibinfo {author} {\bibfnamefont {J.}~\bibnamefont {Goldbeter}}, \
  and\ \bibinfo {author} {\bibfnamefont {J.-M.}\ \bibnamefont {Sparenberg}},\
  }\href {\doibase 10.1103/PhysRevA.65.052710} {\bibfield  {journal} {\bibinfo
  {journal} {Phys. Rev. A}\ }\textbf {\bibinfo {volume} {65}},\ \bibinfo
  {pages} {052710} (\bibinfo {year} {2002})}\BibitemShut {NoStop}%
\bibitem [{\citenamefont {Baye}\ \emph {et~al.}(1998)\citenamefont {Baye},
  \citenamefont {Hesse}, \citenamefont {Sparenberg},\ and\ \citenamefont
  {Vincke}}]{BayeJPB98}%
  \BibitemOpen
  \bibfield  {author} {\bibinfo {author} {\bibfnamefont {D.}~\bibnamefont
  {Baye}}, \bibinfo {author} {\bibfnamefont {M.}~\bibnamefont {Hesse}},
  \bibinfo {author} {\bibfnamefont {J.-M.}\ \bibnamefont {Sparenberg}}, \ and\
  \bibinfo {author} {\bibfnamefont {M.}~\bibnamefont {Vincke}},\ }\href
  {http://stacks.iop.org/0953-4075/31/i=15/a=015} {\bibfield  {journal}
  {\bibinfo  {journal} {J. Phys. B: At. Mol. Opt. Phys.}\ }\textbf {\bibinfo
  {volume} {31}},\ \bibinfo {pages} {3439} (\bibinfo {year}
  {1998})}\BibitemShut {NoStop}%
\bibitem [{\citenamefont {Hesse}\ \emph {et~al.}(2002)\citenamefont {Hesse},
  \citenamefont {Roland},\ and\ \citenamefont {Baye}}]{Hesse2002184}%
  \BibitemOpen
  \bibfield  {author} {\bibinfo {author} {\bibfnamefont {M.}~\bibnamefont
  {Hesse}}, \bibinfo {author} {\bibfnamefont {J.}~\bibnamefont {Roland}}, \
  and\ \bibinfo {author} {\bibfnamefont {D.}~\bibnamefont {Baye}},\ }\href@noop
  {} {\bibfield  {journal} {\bibinfo  {journal} {Nucl. Phys.}\ }\textbf
  {\bibinfo {volume} {A709}},\ \bibinfo {pages} {184} (\bibinfo {year}
  {2002})}\BibitemShut {NoStop}%
\bibitem [{\citenamefont {Hesse}\ \emph {et~al.}(1998)\citenamefont {Hesse},
  \citenamefont {Sparenberg}, \citenamefont {Raemdonck},\ and\ \citenamefont
  {Baye}}]{Hesse199837}%
  \BibitemOpen
  \bibfield  {author} {\bibinfo {author} {\bibfnamefont {M.}~\bibnamefont
  {Hesse}}, \bibinfo {author} {\bibfnamefont {J.-M.}\ \bibnamefont
  {Sparenberg}}, \bibinfo {author} {\bibfnamefont {F.~V.}\ \bibnamefont
  {Raemdonck}}, \ and\ \bibinfo {author} {\bibfnamefont {D.}~\bibnamefont
  {Baye}},\ }\href@noop {} {\bibfield  {journal} {\bibinfo  {journal} {Nucl.
  Phys.}\ }\textbf {\bibinfo {volume} {A640}},\ \bibinfo {pages} {37} (\bibinfo
  {year} {1998})}\BibitemShut {NoStop}%
\bibitem [{\citenamefont {Tanihata}(1996)}]{Tanihata:1995yv}%
  \BibitemOpen
  \bibfield  {author} {\bibinfo {author} {\bibfnamefont {I.}~\bibnamefont
  {Tanihata}},\ }\href {\doibase 10.1088/0954-3899/22/2/004} {\bibfield
  {journal} {\bibinfo  {journal} {J. Phys.}\ }\textbf {\bibinfo {volume}
  {G22}},\ \bibinfo {pages} {157} (\bibinfo {year} {1996})}\BibitemShut
  {NoStop}%
\bibitem [{\citenamefont {Tanihata}\ \emph {et~al.}(1985)\citenamefont
  {Tanihata}, \citenamefont {Hamagaki}, \citenamefont {Hashimoto},
  \citenamefont {Shida}, \citenamefont {Yoshikawa}, \citenamefont {Sugimoto},
  \citenamefont {Yamakawa}, \citenamefont {Kobayashi},\ and\ \citenamefont
  {Takahashi}}]{PhysRevLett.55.2676}%
  \BibitemOpen
  \bibfield  {author} {\bibinfo {author} {\bibfnamefont {I.}~\bibnamefont
  {Tanihata}}, \bibinfo {author} {\bibfnamefont {H.}~\bibnamefont {Hamagaki}},
  \bibinfo {author} {\bibfnamefont {O.}~\bibnamefont {Hashimoto}}, \bibinfo
  {author} {\bibfnamefont {Y.}~\bibnamefont {Shida}}, \bibinfo {author}
  {\bibfnamefont {N.}~\bibnamefont {Yoshikawa}}, \bibinfo {author}
  {\bibfnamefont {K.}~\bibnamefont {Sugimoto}}, \bibinfo {author}
  {\bibfnamefont {O.}~\bibnamefont {Yamakawa}}, \bibinfo {author}
  {\bibfnamefont {T.}~\bibnamefont {Kobayashi}}, \ and\ \bibinfo {author}
  {\bibfnamefont {N.}~\bibnamefont {Takahashi}},\ }\href {\doibase
  10.1103/PhysRevLett.55.2676} {\bibfield  {journal} {\bibinfo  {journal}
  {Phys. Rev. Lett.}\ }\textbf {\bibinfo {volume} {55}},\ \bibinfo {pages}
  {2676} (\bibinfo {year} {1985})}\BibitemShut {NoStop}%
\bibitem [{\citenamefont {Kukulin}\ \emph {et~al.}(1986)\citenamefont
  {Kukulin}, \citenamefont {Krasnopolsky}, \citenamefont {Voronchev},\ and\
  \citenamefont {Sazonov}}]{kukulin86}%
  \BibitemOpen
  \bibfield  {author} {\bibinfo {author} {\bibfnamefont {V.}~\bibnamefont
  {Kukulin}}, \bibinfo {author} {\bibfnamefont {V.}~\bibnamefont
  {Krasnopolsky}}, \bibinfo {author} {\bibfnamefont {V.}~\bibnamefont
  {Voronchev}}, \ and\ \bibinfo {author} {\bibfnamefont {P.}~\bibnamefont
  {Sazonov}},\ }\href@noop {} {\bibfield  {journal} {\bibinfo  {journal} {Nucl.
  Phys.}\ }\textbf {\bibinfo {volume} {A453}},\ \bibinfo {pages} {365}
  (\bibinfo {year} {1986})}\BibitemShut {NoStop}%
\bibitem [{\citenamefont {Zhukov}\ \emph {et~al.}(1993)\citenamefont {Zhukov},
  \citenamefont {Danilin}, \citenamefont {Fedorov}, \citenamefont {Bang},
  \citenamefont {Thompson},\ and\ \citenamefont {Vaagen}}]{Zhukov1993151}%
  \BibitemOpen
  \bibfield  {author} {\bibinfo {author} {\bibfnamefont {M.}~\bibnamefont
  {Zhukov}}, \bibinfo {author} {\bibfnamefont {B.}~\bibnamefont {Danilin}},
  \bibinfo {author} {\bibfnamefont {D.}~\bibnamefont {Fedorov}}, \bibinfo
  {author} {\bibfnamefont {J.}~\bibnamefont {Bang}}, \bibinfo {author}
  {\bibfnamefont {I.}~\bibnamefont {Thompson}}, \ and\ \bibinfo {author}
  {\bibfnamefont {J.}~\bibnamefont {Vaagen}},\ }\href {\doibase
  10.1016/0370-1573(93)90141-Y} {\bibfield  {journal} {\bibinfo  {journal}
  {Phys. Rep.}\ }\textbf {\bibinfo {volume} {231}},\ \bibinfo {pages} {151 }
  (\bibinfo {year} {1993})}\BibitemShut {NoStop}%
\bibitem [{\citenamefont {Descouvevemont}\ \emph {et~al.}(2012)\citenamefont
  {Descouvevemont}, \citenamefont {Pinilla},\ and\ \citenamefont
  {Baye}}]{Descou12}%
  \BibitemOpen
  \bibfield  {author} {\bibinfo {author} {\bibfnamefont {P.}~\bibnamefont
  {Descouvevemont}}, \bibinfo {author} {\bibfnamefont {E.}~\bibnamefont
  {Pinilla}}, \ and\ \bibinfo {author} {\bibfnamefont {D.}~\bibnamefont
  {Baye}},\ }\href@noop {} {\bibfield  {journal} {\bibinfo  {journal} {Prog.
  Theor. Phys. (Supp.)}\ }\textbf {\bibinfo {volume} {196}},\ \bibinfo {pages}
  {1} (\bibinfo {year} {2012})}\BibitemShut {NoStop}%
\bibitem [{\citenamefont {Bacca}\ \emph {et~al.}(2009)\citenamefont {Bacca},
  \citenamefont {Schwenk}, \citenamefont {Hagen},\ and\ \citenamefont
  {Papenbrock}}]{Bacca:2009yk}%
  \BibitemOpen
  \bibfield  {author} {\bibinfo {author} {\bibfnamefont {S.}~\bibnamefont
  {Bacca}}, \bibinfo {author} {\bibfnamefont {A.}~\bibnamefont {Schwenk}},
  \bibinfo {author} {\bibfnamefont {G.}~\bibnamefont {Hagen}}, \ and\ \bibinfo
  {author} {\bibfnamefont {T.}~\bibnamefont {Papenbrock}},\ }\href {\doibase
  10.1140/epja/i2009-10815-5} {\bibfield  {journal} {\bibinfo  {journal} {Eur.
  Phys. J.}\ }\textbf {\bibinfo {volume} {A42}},\ \bibinfo {pages} {553}
  (\bibinfo {year} {2009})}\BibitemShut {NoStop}%
\bibitem [{\citenamefont {Brida}\ and\ \citenamefont
  {Nunes}(2010)}]{Brida:2010ae}%
  \BibitemOpen
  \bibfield  {author} {\bibinfo {author} {\bibfnamefont {I.}~\bibnamefont
  {Brida}}\ and\ \bibinfo {author} {\bibfnamefont {F.}~\bibnamefont {Nunes}},\
  }\href {\doibase 10.1016/j.nuclphysa.2010.06.012} {\bibfield  {journal}
  {\bibinfo  {journal} {Nucl. Phys.}\ }\textbf {\bibinfo {volume} {A847}},\
  \bibinfo {pages} {1} (\bibinfo {year} {2010})}\BibitemShut {NoStop}%
\bibitem [{\citenamefont {Navr\'atil}\ and\ \citenamefont
  {Ormand}(2003)}]{PhysRevC.68.034305}%
  \BibitemOpen
  \bibfield  {author} {\bibinfo {author} {\bibfnamefont {P.}~\bibnamefont
  {Navr\'atil}}\ and\ \bibinfo {author} {\bibfnamefont {W.~E.}\ \bibnamefont
  {Ormand}},\ }\href {\doibase 10.1103/PhysRevC.68.034305} {\bibfield
  {journal} {\bibinfo  {journal} {Phys. Rev. C}\ }\textbf {\bibinfo {volume}
  {68}},\ \bibinfo {pages} {034305} (\bibinfo {year} {2003})}\BibitemShut
  {NoStop}%
\bibitem [{\citenamefont {Pieper}\ and\ \citenamefont
  {Wiringa}(2001)}]{Pieper2001}%
  \BibitemOpen
  \bibfield  {author} {\bibinfo {author} {\bibfnamefont {S.}~\bibnamefont
  {Pieper}}\ and\ \bibinfo {author} {\bibfnamefont {R.}~\bibnamefont
  {Wiringa}},\ }\href {\doibase 10.1146/annurev.nucl.51.101701.132506}
  {\bibfield  {journal} {\bibinfo  {journal} {Annu. Rev. Nucl. Part. Sci.}\
  }\textbf {\bibinfo {volume} {51}},\ \bibinfo {pages} {53} (\bibinfo {year}
  {2001})}\BibitemShut {NoStop}%
\bibitem [{\citenamefont {Brodeur}\ \emph {et~al.}(2012)\citenamefont
  {Brodeur}, \citenamefont {Brunner}, \citenamefont {Champagne}, \citenamefont
  {Ettenauer}, \citenamefont {Smith} \emph {et~al.}}]{Brodeur:2012zz}%
  \BibitemOpen
  \bibfield  {author} {\bibinfo {author} {\bibfnamefont {M.}~\bibnamefont
  {Brodeur}}, \bibinfo {author} {\bibfnamefont {T.}~\bibnamefont {Brunner}},
  \bibinfo {author} {\bibfnamefont {C.}~\bibnamefont {Champagne}}, \bibinfo
  {author} {\bibfnamefont {S.}~\bibnamefont {Ettenauer}}, \bibinfo {author}
  {\bibfnamefont {M.}~\bibnamefont {Smith}},  \emph {et~al.},\ }\href {\doibase
  10.1103/PhysRevLett.108.052504} {\bibfield  {journal} {\bibinfo  {journal}
  {Phys. Rev. Lett.}\ }\textbf {\bibinfo {volume} {108}},\ \bibinfo {pages}
  {052504} (\bibinfo {year} {2012})}\BibitemShut {NoStop}%
\bibitem [{\citenamefont {Varshalovich}\ \emph {et~al.}(1988)\citenamefont
  {Varshalovich}, \citenamefont {Moskalev},\ and\ \citenamefont
  {Khersonski{\u\i}}}]{varshalovich}%
  \BibitemOpen
  \bibfield  {author} {\bibinfo {author} {\bibfnamefont {D.}~\bibnamefont
  {Varshalovich}}, \bibinfo {author} {\bibfnamefont {A.}~\bibnamefont
  {Moskalev}}, \ and\ \bibinfo {author} {\bibfnamefont {V.}~\bibnamefont
  {Khersonski{\u\i}}},\ }\href@noop {} {\emph {\bibinfo {title} {Quantum Theory
  of Angular Momentum}}}\ (\bibinfo  {publisher} {World Scientific Publishing
  Company, Incorporated},\ \bibinfo {year} {1988})\BibitemShut {NoStop}%
\end{thebibliography}

%

\end{document}